\documentclass{article}

\usepackage{VMpreamble}
\usepackage{TikzPictures}
\usepackage{jheppub}

\newcommand{\LLL}{\hat{\mathcal{L}}}
\newcommand{\LL}{\hat{L}}
\tikzset{
    externalline/.style={-{Circle[open,length=3pt,width=3pt]}}
}
\usepackage{upgreek}

\graphicspath{ {./Pictures/} } 

 \newcommand{\beq}{\begin{equation}}
   \newcommand{\eeq}{\end{equation}}
\newcommand\beqa{\begin{eqnarray}}
    \newcommand\eeqa{\end{eqnarray}}
\def\<{\left<}
\def\>{\right>}
\def\d{\partial}
\newcommand{\nn}{\nonumber}
\newcommand{\eq}[1]{(\ref{#1})}

\title{Yangian symmetry, GKZ equations and integrable Feynman graphs in conformal variables

}

\emailAdd{fedor.levkovich$\bullet$gmail.com}
\emailAdd{mishnyakovvv$\bullet$gmail.com}

\author[a]{Fedor Levkovich-Maslyuk,}
\author[b]{Victor Mishnyakov}

\affiliation[a]{Centre for Mathematical Science, City St George's, University of London, Northampton Square, EC1V 0HB, London, UK
 } 

\affiliation[b]{Nordita, KTH Royal Institute of Technology and Stockholm University, 
Hannes Alfv\'ens v\"ag 12, SE-106 91 Stockholm, Sweden  }

\abstract{
We study the differential equations that follow from Yangian symmetry which was recently observed for a large class of conformal Feynman graphs, originating from integrable `fishnet' theories. We derive, for the first time, the explicit general form of these equations in the most useful conformal cross-ratio variables, valid for any spacetime dimension. This allows us to explore their properties in detail. In particular, we observe that for general Feynman graphs a large set of terms in the Yangian equations can be identified with famous GKZ (Gelfand-Kapranov-Zelevinsky) hypergeometric operators. We also show that for certain nontrivial graphs the relation with GKZ systems is exact, opening the way to using new powerful solution methods. As a side result, we also elucidate the constraints on the topology and parameter space of Feynman graphs stemming from Yangian invariance.

}

\begin{document}

\maketitle

\section{Introduction}

Recent years have seen a fruitful interaction between two major directions in modern QFT. The first one is the study of multi-loop Feynman integrals, which has been rather extensively developing in recent years with progress both on the formal and application side of the problem \cite{Weinzierl:2022eaz,Vanhove:2018mto,Abreu:2022mfk,Blumlein:2022qci,Bourjaily:2022bwx}. The second one is exploration of integrability and in particular Yangian symmetry for quantum field theories. Namely, integrable techniques, long known from spin-chains and integrable 2d field theories, have been very successfully applied to $N=4$ SYM theory in four dimensions \cite{Beisert:2010jr} as well as to other higher-dimensional models. The combination of the two approaches often leads to remarkable new insights for specific Feynman graphs, and this is the subject we will explore in this paper.

We will focus on the large class of conformal Feynman integrals that originate from models known as fishnet theories. 
First studied by Zamolodchikov \cite{Zamolodchikov:1980mb}, fishnet Feynman graphs were recently found to appear in a certain limit of $N=4$ SYM theory giving the fishnet CFT model  \cite{Gurdogan:2015csr}.  One of their integrable properties manifests itself as a symmetry under the conformal Yangian algebra \cite{Chicherin:2017cns,Chicherin:2017frs} that has led to new remarkable insights as well as explicit calculation of a number of nontrivial Feynman integrals \cite{Loebbert:2024qbw,
Duhr:2024hjf,
Loebbert:2024fsj,
Duhr:2023eld,
Loebbert:2022nfu,
Duhr:2022pch,
Corcoran:2021gda,
Loebbert:2021qef,
Corcoran:2020epz,
Loebbert:2020glj,
Loebbert:2020tje,
Loebbert:2020hxk,
Loebbert:2019vcj}, see also the review \cite{Chicherin:2022nqq}\footnote{Let us also note that other applications of Yangian-type symmetries have been also explored before, especially in the context of N=4 SYM, see e.g. \cite{Drummond:2009fd,Drummond:2008vq,Beisert:2010gn,Arkani-Hamed:2012zlh,Huang:2010qy,Bargheer:2010hn,Beisert:2017pnr}.}. This symmetry appears to be highly constraining and often allows one to calculate the Feynman integrals when supplemented with prescribed symmetries and boundary conditions \cite{Loebbert:2019vcj}. Furthermore, the original fishnet graphs are not the most general ones that have the Yangian symmetry property. A much larger class of fishnet-type graphs were recently proved to be Yangian invariant in \cite{Kazakov:2023nyu}. They are also ultimately based on Zamolodchikov's construction and appear in generalised versions of the fishnet theory known as Loom CFTs \cite{Kazakov:2022dbd}. Fishnet graphs  were also studied by various other integrability methods \cite{Kazakov:2018gcy,Derkachov:2018rot,Alfimov:2023vev}.

The Yangian symmetry can be realized in the form of a system of differential equations, which is how it has been utilized in the bootstrap approach \cite{Loebbert:2019vcj}. More generally, various approaches based on differential equations have proven very useful for the study of Feynman integrals. For  complicated graphs with multiple parameters like masses and momenta one typically uses the differential equations originating from the IBP relations \cite{Bezuglov:2020tff, Kotikov:2021tai,Abreu:2020jxa}. On the other hand, a more geometric approach produces the so-called Picard-Fuchs equations, which are differential equations for periods of the corresponding manifolds. The most studied case are the banana graphs and their generalizations like the ice-cone family \cite{Bonisch:2021yfw,Lairez:2022zkj,delaCruz:2024xit}. For these graphs, one finds that the underlying geometry is that of a Calabi-Yau manifold, which is also reflected in the nature of the Picard-Fuchs equations. Curiously, it has been recently discovered that the fishnet integrals in two dimensions are in fact also periods of Calabi-Yau manifolds, while the Yangian differential equations are nothing but the corresponding Picard-Fuchs equations \cite{Duhr:2022pch}. This remarkable additional geometric structure has already been used to calculate a range of new fishnet integrals \cite{Duhr:2024hjf}.

Another important type of differential equations that arise for Feynman integrals are the Gel'fand-Kapranov-Zelevinsky equations, or GKZ systems \cite{gel1989hypergeometric,Gelfand:1990bua}. 
In fact, they appear in a range of contexts in mathematics and physics, from solving roots of generic algebraic equations \cite{sturmfels2000solving}   and non-gaussian integrals \cite{Morozov:2009fc} to mirror symmetry \cite{Hosono:1995bm,batyrev1993variations}. It was also quite recently found that (as anticipated already in \cite{gel1989hypergeometric}) many Feynman integrals satisfy GKZ equations \cite{nasrollahpoursamami2016periods,delaCruz:2019skx, Vanhove:2018mto,Grimm:2024tbg}, see also \cite{Pal:2023kgu, Rigatos:2022eos} for further results in a direction related to what we explore in this paper. Practical importance of GKZ systems for us stems from the fact that there exists a powerful and well-developed solution theory for them. In particular, under very general assumptions one can typically write the solution in terms of ${\cal A}$-hypergeometric functions whose properties are well controlled. Thus establishing a link with GKZ theory can open the way to explicitly computing new classes of Feynman integrals, or at least finding an explicit basis of special functions whose linear combinations give the integral.

At this point let us summarise the main ideas and applications that motivate the study of Yangian invariant Feynman integrals: 
\begin{itemize}
    \item The most apparent one, as explained above, is the calculation of Feynman integrals themselves, which is often made possible by the use of Yangian symmetry.

    \item In the context of the study of differential equations, Yangian symmetry provides a rather interesting case where the differential operators  form a well-known algebra. Even more curiously, the algebra is only clearly seen in certain variables. Namely, whereas for practical computations one rewrites the Yangian equations in conformal cross-ratios, the algebra itself is only apparent in the variables corresponding to position-space coordinates of external legs. 
    \item As discussed above, the Feynman integrals in question correspond to special large $N$ CFTs, where they make up the perturbative expansion. Thus Yangian symmetry offers the prospect of calculating many of these graphs and the corresponding observables in these fishnet theories. One may even speculate that perhaps this could eventually be possible for all these graphs by virtue of integrability, leading to a complete perturbative solution of these models.

\end{itemize}

In this paper, we study Yangian differential equations for Feynman integrals in high generality.  Apart from the case of $D=2$ spacetime dimensions, so far the properties of these equations have been closely studied only for several specific graphs -- the 4-point and 6-point cross, and the 6-point double-cross \cite{Loebbert:2019vcj}. Several more general statements were also put forward regarding the structure of the equations and the number of independent constraints. Here we address some of these and related questions in the rather general situation. Below we highlight our results.

\bigskip 
\noindent\textbf{Summary of main results.} One of our main results concerns the explicit form of the Yangian differential equations in conformal variables.   Conformal symmetry imposes that the Feynman integral should be a function of conformal cross-ratios, yet Yangian symmetry is formulated in the original position coordinates of the external legs. It is far from obvious in general how to concisely rewrite the Yangian differential operators in cross-ratio variables, yet it is precisely this form which is needed for the most important applications. Without even the explicit form of the equations it is difficult to establish their general properties and describe their space of solutions. In this paper we show how to obtain the general form of the equations in the cross-ratios (under certain assumptions) for any dimension, any graph, and any choice of cross-ratio basis. Explicitly, we find that the resulting Yangian equations have a rather compact form:
\begin{equation}\label{eq:PDEik}
\begin{split}
        \PDE_{ik} &= 2\left( \sum_{l>j>i} - \sum_{l<j<i} + \sum_{l<k<i ; j} - \sum_{l>k>i; j } \right) \chi_{iklj} \theta_{il}\theta_{jk} +\sum_{j\neq i}(\delta_{j>i}-\delta_{j<i})\theta_{ik}\theta_{ij} 
        \\
&\;+\left(\delta_{i<k}\left(\Delta_k-D \right)- \delta_{i>k}\left(\Delta_i-D \right)  \right)\theta_{ik} + 2(s_i-s_k)\theta_{ik}
\end{split}
\end{equation}
Here the indices range over $1,\dots,N$ where $N$ is the number of external points, and $s_i$ are the evaluation parameters of the Yangian, while $\theta_{ij}$ are first order differential operators in the cross-ratios $\xi^A$,
\beq
    \theta_{ij}=\sum_A\alpha_{ij}^A\xi^A\frac{\d}{\d\xi^A}+\beta_{ij}
\eeq
with $\alpha_{ij}, \; \beta_{ij}$ defined by the choice of conformal parameterisation of the external points, and $\chi_{ijkl}$ are standard 4-point cross-ratios (see section \ref{sec:crr} for details). These equations match the specific cases studied in the literature for particular graphs, and extend them to the general situation.

    Having obtained the general form of the equations we can now study them in-depth. In particular we address the question of the precise relation between Yangian and GKZ equations. The particular case of the $N$-cross integral was shown in \cite{Pal:2023kgu} to satisfy a GKZ system. Here we investigate the links with GKZ equations for more general graphs. First, we find that the Yangian equations in general can be recast as a sum of GKZ  differential operators (which encapsulate the most involved parts of the equation) plus a remainder term built up only from 1st order derivatives. Furthermore, we study the conditions under which the remainder term vanishes and the Yangian precisely reduces to a GKZ system. We find that this happens for other examples in addition to the cross integrals discussed in \cite{Pal:2023kgu}, offering new ways to try to constrain or even fully compute novel Feynman integrals. In particular we find that, nontrivially,  some integrals should be expressed as linear combinations of the same hypergeometric functions as the multi-leg cross integral. We also present indirect arguments regarding the structure of the equations, suggesting a wider range of applications to be explored in the future.

Importantly,  in contrast to the standard approach via the Lee-Pomeransky representation, we do not need to enlarge the parameter space of the integral and then only later reduce it in order to establish a link with GKZ systems. Instead we find that GKZ equations are written directly in terms of the simple coordinate variables parameterising the positions of external legs of the graph.

As an important side question, we also investigate which types of graphs can be Yangian invariant. In principle, the Loom construction gives the answer. However, in some cases it is too restrictive, and furthermore one may still ask what kind of graph topologies are allowed, as well as what are the precise constraints on the graph's parameters. We address this set of questions and in some cases find concise descriptions for these constraints.

\bigskip 
\noindent\textbf{Structure of the paper.} This paper is organised as follows. In section 2 we review the main ideas of Yangian symmetry and the Loom construction of integrable Feynman graphs. In section 3 we study constraints on the graph geometry and its parameters imposed by Yangian invariance in general. In section 4 we present one of our main results, deriving the compact and explicit form of Yangian equations in conformal cross-ratio variables. We also elucidate the nontrivial requirements on the graph parameters following from self-consistency of the equations. In section 5 we discuss the relation of the Yangian with GKZ systems, and finally in section 6 we discuss future directions. The appendices contain a number of more technical details.

\section{Review of Yangian symmetry for Loom graphs}

In this section we briefly review the concept of Yangian symmetry and the Loom construction from \cite{Kazakov:2022dbd,Kazakov:2023nyu}. We define all the necessary differential equations and prescriptions for the evaluation parameters in  Yangian equations.

\subsection{Conformal Yangian symmetry}
Throughout this paper we consider massless Feynman integrals, that is, to a graph $\Gamma$ one associates the integral:
\begin{equation}
    I_\Gamma(\mathbf{x}|\pmb{\Delta}) = \int \prod_{k \in \text{internal}} d^D x_{k} \dfrac{1}{\prod\limits_{(i,j) \in \text{edges}} x_{ij}^{2\Delta_{ij}} }
\end{equation}
where $\mathbf{x} = (x_1,\ldots x_N)$ are the positions of external legs in $D$-dimensional space,  $x_k \in \mathbb{R}^D$, and $\pmb{\Delta}=
\{
\Delta_1,\dots,\Delta_K\}$ is a collection of propagators powers (we also call them dimensions) on which the integral depends. For our purposes, the signature of space-time is not crucial, even though one 
has to take it into account in other scenarios \cite{Corcoran:2020epz}. The dimension $D$ is also a parameter of this function, but  we will not list it explicitly.  In this paper we will only consider  conformal Feynman integrals. This requires that the sum of dimensions in each internal (integrated) vertex equals to $D$. Therefore, the number of independent dimensions $\Delta_j$ is equal to the total number of edges minus the number of internal vertices.

Conformal $\mathfrak{so}(D,2)$ symmetry is realized with the following differential operator representation:
\begin{equation}
\label{cop}
\begin{gathered}
            P_{\mu,j}=-i \dfrac{\partial}{\partial x^\mu_j} \equiv\hat p_\mu\ , \ \ D_j=x^\mu \hat p_\mu-i\Delta_j \ ,
\\
    L_{j}^{\mu\nu}=x_j^\nu\hat p_j^\mu-x_j^\mu\hat p_j^\nu \equiv \hat\ell_{\mu\nu}\ , \ \ \  K_{\mu,j}=2x^\nu\hat\ell_{\nu\mu}+x_\nu x^\nu\hat p_\mu-2i\Delta_j x^\mu_j \ .
\end{gathered}
\end{equation}
The generators act on the coordinate of the $j$'th external point of the integral. The conformal dimension associated to an external vertex is the total dimension of all the propagators that connect to this vertex. Conformal invariance of the  integral is then realized by acting on all the external legs:
\begin{equation}\label{eq:ConformalSymmetryMomentum}
    P^\mu \, I_\Gamma(\mathbf{x}|\pmb{\Delta}) = \sum_{j=1}^N P_j^\mu \, I_\Gamma(\mathbf{x}|\pmb{\Delta}) = 0
\end{equation}
and similarly for other generators. Conformal symmetry is quite a powerful constraint as it implies that the integrals depend only on conformal invariants, and allows using such tools as the star-triangle transformation (see Appendix \ref{sec:AppendixDoubleStarIntegral}). The condition that the sum of propagator powers is $D$ at each internal vertex also implies that all the integrals converge, at least in terms of  power counting. Conformal symmetry is sufficient to fix the form of two and three point integrals. However, at four points (for $D>1$), non-trivial conformal invariants appear. In that case conformal symmetry only fixes the integral up to an arbitrary function of conformal invariants as we discuss in more detail in section \ref{sec:crr}.

As we mentioned in the Introduction, for certain types of graphs there exists an additional symmetry presented in the form of the Yangian algebra of the conformal group. There are quite a few definitions of Yangians \cite{Loebbert:2016cdm}. The one that we use here is realized in terms of the Lie algebra $\mathfrak{g}$ generators $J^{a}_i$ which act at a given vertex $i$. The Yangian $Y(\mathfrak{g})$ is generated by the level-zero generators:
\begin{equation}
    J^a = \sum_i J^{a}_i \,, 
\end{equation}
which in our case are the ones in equation \eqref{eq:ConformalSymmetryMomentum}, and the level-one generators:
\begin{equation}
    \hat{J}^{a} = \sum_{i>j} f^{a}_{\phantom{a}bc} J^b_i J^c_j+ \sum_i s_i J_i^{a}
\end{equation}
where $s_i$ are known as the evaluation parameters. We will be interested in Feynman integrals invariant under the Yangian, i.e. annihilated by the level-zero and level-one generators.  Note that the commutation relations
\begin{equation}
    [J^a,\hat{J}^b] =f^{ab}_{\phantom{ab}c} \hat{J}^{c}
\end{equation}
imply that for conformal Feynman integrals it is enough to impose invariance under only one of the level-one generators. For the conformal algebra we typically work with the level-one momentum generator in this role, which is given by \cite{Chicherin:2017cns}: 
\beq
\label{ph}
    {\widehat P}^\mu=-\frac{i}{2}\sum_{j<k}[(L_j^{\mu\nu}+g^{\mu\nu}D_j)P_{k,\nu}-(j\leftrightarrow k)]+\sum_i s_i P_i^\mu
\eeq
Then Yangian invariance of the integral is the following constraint:
\begin{equation}
     {\widehat P}^\mu \, I_\Gamma(\mathbf{x}|\pmb{\Delta}) = 0
\end{equation}
The evaluation parameters actually depend on the graph and are different in each case. They are expressed linearly in $\Delta_i$'s in ways depending on the graph topology. Fixing these parameters is quite a complicated task, which was solved in general in \cite{Kazakov:2023nyu}. We will review the resulting prescription below. Note as well that the evaluation parameters are only determined up to an overall constant shift $s_i  \rightarrow s_i+c$, since it just represents adding a level-one generator that itself annihilates the integral. Thus, one can always set e.g. $s_1=0$.

\subsection{The Loom construction}

 \begin{figure}[t]
 \centering
   \includegraphics[scale=0.55]{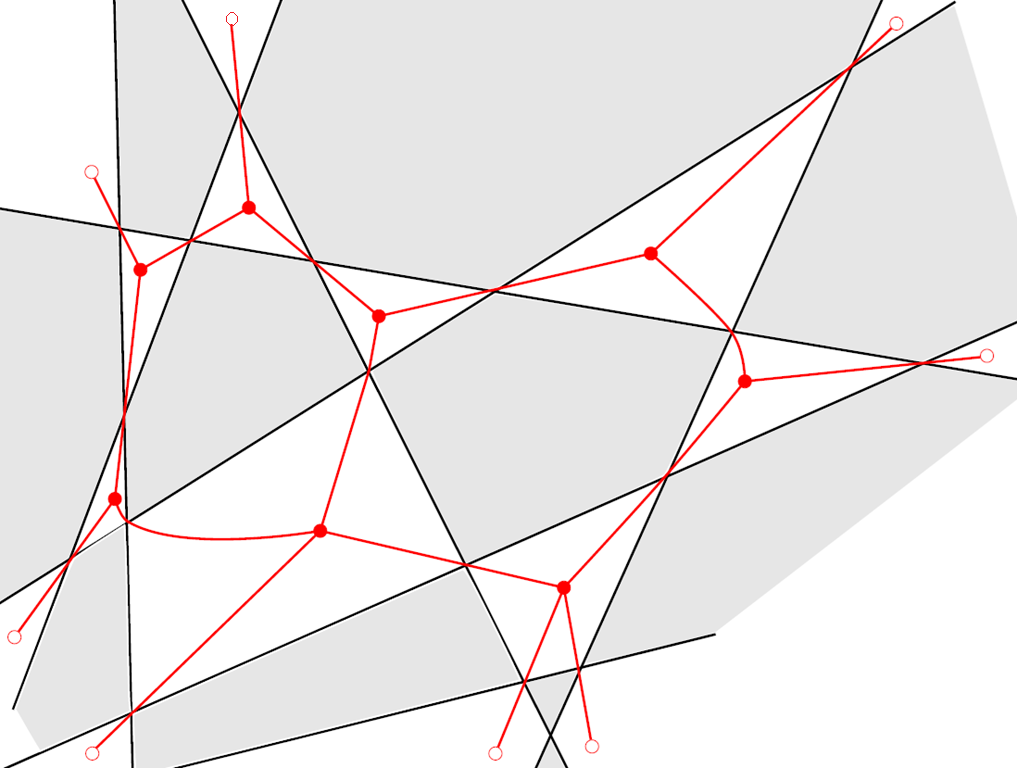}  \caption{ An example of a Feynman graph created using the Loom construction \cite{Kazakov:2022dbd,Kazakov:2023nyu}. Propagators are depicted as red lines, internal vertices as filled red circles, and external points as blank red circles. The black lines represent the original Baxter lattice. Each internal vertex and external point is assigned a $D$-dimensional coordinate, with integration performed over the coordinates of internal points. }

   \label{fig:loomexnolass}
 \end{figure}

\begin{figure}[t]
 \centering
\includegraphics[scale=0.455]{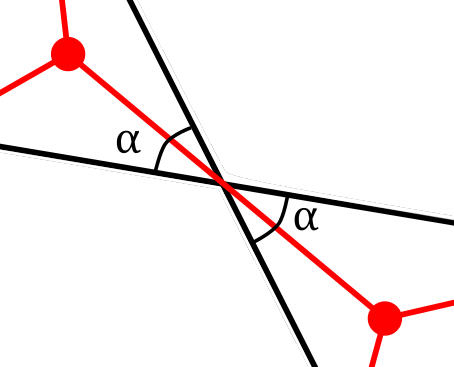}  \caption{A propagator that passes through a vertex with angle $\alpha$ between the Baxter lattice lines carries the conformal dimension defined by \eq{da}.}\label{fig:propdem1}
\end{figure}

The large class of Yangian invariant  graphs that were described in \cite{Kazakov:2022dbd} and \cite{Kazakov:2023nyu} are constructed geometrically -- an example is given on figure \ref{fig:loomexnolass}. The starting point is a Baxter lattice -- namely, a finite set of lines on the plane. Some of these lines may be parallel to each other, but we do not allow triple or higher intersections. The Baxter lattice divides the plane into a number of faces (polygons) which admits a checkerboard-type coloring. On the diagrams we will draw these faces as white or grey. The Feynman diagram itself will be drawn on the graph which is the \textit{dual}\footnote{i.e. its vertices correspond to faces of the original graph, and vice versa} to the set of the white faces. The Feynman integral is then constructed as follows:
 \begin{itemize}
     \item 
    For any of the white faces which is a closed polygon,  we may place an internal vertex of the Feynman graph inside it. If we do that, we draw propagators that go out of this vertex to each of its neighboring white faces, passing through the angles of the polygon.  Then to every one of the internal vertices we associate a $D$-dimensional coordinate that will be integrated over. 
     \item A given propagator that comes out of an internal vertex can be either an external leg of the diagram, or else it can connect to the internal vertex located in the adjacent white face.
     \item The propagator connecting two points $x_1$ and $x_2$ reads
   \beq
   \label{cprop}
     \frac{1}{|x_{1}-x_{2}|^{2\Delta}}\ ,
   \eeq
    Importantly, the scaling dimension $\Delta$ here is fixed by the angle of the polygon through which the propagator goes, according to  (see figure \ref{fig:propdem1})
   \beq
   \label{da}
    \Delta=D\frac{\pi-\alpha}{2\pi} \ .
 \eeq
 \end{itemize}

This gives as a result a Feynman diagram that defines an integral in coordinate space, with propagators having the form \eq{cprop}.

This construction ensures that the resulting integral is always conformal. The sum of scaling dimensions for propagators at each vertex is $D$ due to the  geometric constraint on the sum of the angles of any closed polygon. To see this, notice that $\pi -\alpha$ in \eq{da} is the external angle at the vertex of the polygon, and the sum of all such angles for an $n$-gon is $\pi n - \pi(n-2)=2\pi$ (regardless of $n$). 
\\

The proof of Yangian invariance of such graphs in \cite{Kazakov:2023nyu} utilized the Lasso method. The essential ingredient is the transfer matrix -- a product of Lax operators of the conformal group. On the one hand it has the integral as its eigenstate, and on the other hand it is a generating function of Yangian operators. The specific details of this technique are not important here. The main point   are the prescriptions for the shifts  $\delta_i^{+}$ and $\delta_i^{-}$ in the Lax operators. These are  combinations of the dimensions of external legs calculated according to certain rules. One starts with any leg, and then goes along the graph, say, clockwise, and calculates the shifts for each leg, depending on whether the previous one connected to the same internal vertex or not. 
In general, going from vertex $i$ to vertex $i+1$ the shifts are found as \cite{Kazakov:2023nyu} (see also \cite{Loebbert:2019vcj,  Loebbert:2020hxk} for related examples in special cases)
\begin{equation}
\begin{split}
        &\delta_{i+1}^+=\delta_i^+ + \Delta_{i+1}+\sum_{j=1}^p \left(\Delta^{(int)}_j - D/2 \right)
        \\
        &\delta_{i+1}^-=\delta_i^-+\Delta_{i}+\sum_{j=1}^p\left(\Delta^{(int)}_j - D/2 \right)
\end{split}
\end{equation}
where $p$ is the number of internal propagators separating the leg $i$ and $i+1$, and $\Delta^{(int)}_{j}$ are their dimensions. When the legs end on the same internal vertex one has $p=0$ and the sum is absent. The initial conditions for the first shift are:
\begin{equation}
    \delta_1^+ =\Delta_1 \, , \quad \delta_1^- =D/2 \,.
\end{equation}
These are illustrated and discussed in detail in \cite{Kazakov:2023nyu}. Finally, once the shifts are calculated, the evaluation parameters are read off using:
\beq
\label{vres}
    s_k=\frac{1}{2}\sum_{j\neq k}(\delta_j^+ + \delta_j^-+D/2) \ .
\eeq

\section{Restrictions on Feynman graphs from Yangian invariance}
In this section we discuss  the kinds of Feynman graph topology and restrictions on the Feynman integral parameters that allow for Yangian invariance a Loom representation.

\subsection{The Loom and dual conformal symmetry}

Let us start by discussing how the Loom construction is related to dual conformal symmetry.

It is generally understood that roughly speaking Yangian integrability should be equivalent to simultaneous conformal and dual conformal symmetry.  Recall that dual conformal symmetry in the context we discuss corresponds to conformal symmetry in momentum rather than position space. More precisely, going to momentum space is also accompanied by replacing the original graph with its dual graph (whose vertices correspond to faces of the original graph, and vice versa), and it is this graph which should be conformal  (we refer e.g. to the review \cite{Chicherin:2022nqq} for details). Note that for intersecting edges of the original and dual graph their conformal dimensions should be related as $\Delta+\Delta_{dual}=D$.

Usual conformal symmetry restricts the sum of dimensions at each \textit{vertex} of the graph to be $D$,
\beq
\label{coc1}
    \Delta_1+\dots+\Delta_M=D
\eeq
Fittingly, one can observe that instead 
\textit{dual} conformal symmetry restricts the sum of dimensions of propagators forming each \textit{face} of the original graph: for a face with $K$ edges it requires
\beq\label{cod1}
    \Delta_1+\dots+\Delta_K=D(K-2)/2
\eeq
This is straightforward to verify for simple examples like the box/cross integrals, and it is likely possible to prove this in general as well.

Remarkably, this constraint \eq{cod1} is automatically incorporated in the Loom construction. Indeed, for any closed polygon the dimensions of the propagators for a Loom graph will satisfy \eq{cod1} due to their geometric relation with the angles \eq{da} as one can easily verify. Thus the Loom graphs should automatically have both usual \textit{and} dual conformal symmetry, serving as another example demonstrating explicitly the essential equivalence between Yangian symmetry, dual conformal symmetry and integrability. 

Another important related question is the following -- given a graph built via the Loom, what are the constraints on the dimensions $\Delta_i$ in it following from the geometry and the relation with the angles \eq{da}? Clearly, among them are the conformal symmetry constraint \eq{coc1} and the dual conformal symmetry relations \eq{cod1} -- but could there be more constraints? After all, the Loom construction involves a whole nontrivial set of angles between lines of the Baxter lattice, which could potentially indirectly relate propagator powers belonging to remote parts of the graph. Indeed in \cite{Kazakov:2023nyu} this possibility was mentioned (under the name of 'non-local constraints') and it has been unclear so far what is the minimal set of constraints. Here we propose a simple answer -- namely, that all geometric constraints amount in the end to imposing only usual conformal symmetry and dual conformal symmetry (in the sense of \eq{cod1}), i.e.:
\beqa
\label{con1}
    &\text{At each vertex the sum of }  \Delta\text{'s is }D.&\\
    \label{con2}
    & \text{Around each face with }K\text{ edges the sum of }\Delta\text{'s is }D(K-2)/2.&
\eeqa
While this observation is rather natural, it clarifies a potentially complicated geometric question. We present a geometric 
argument supporting it in appendix \ref{app:geom}, and it would be also interesting to establish it rigorously\footnote{Potentially the set of constraints could also depend on the way a particular is drawn on the Loom as discussed in \cite{Kazakov:2023nyu}. We expect that the constraints we propose here correspond to the minimal set of requirements among different ways of drawing the graph.}.

\begin{figure}[h]
 \centering
\includegraphics[scale=0.155]{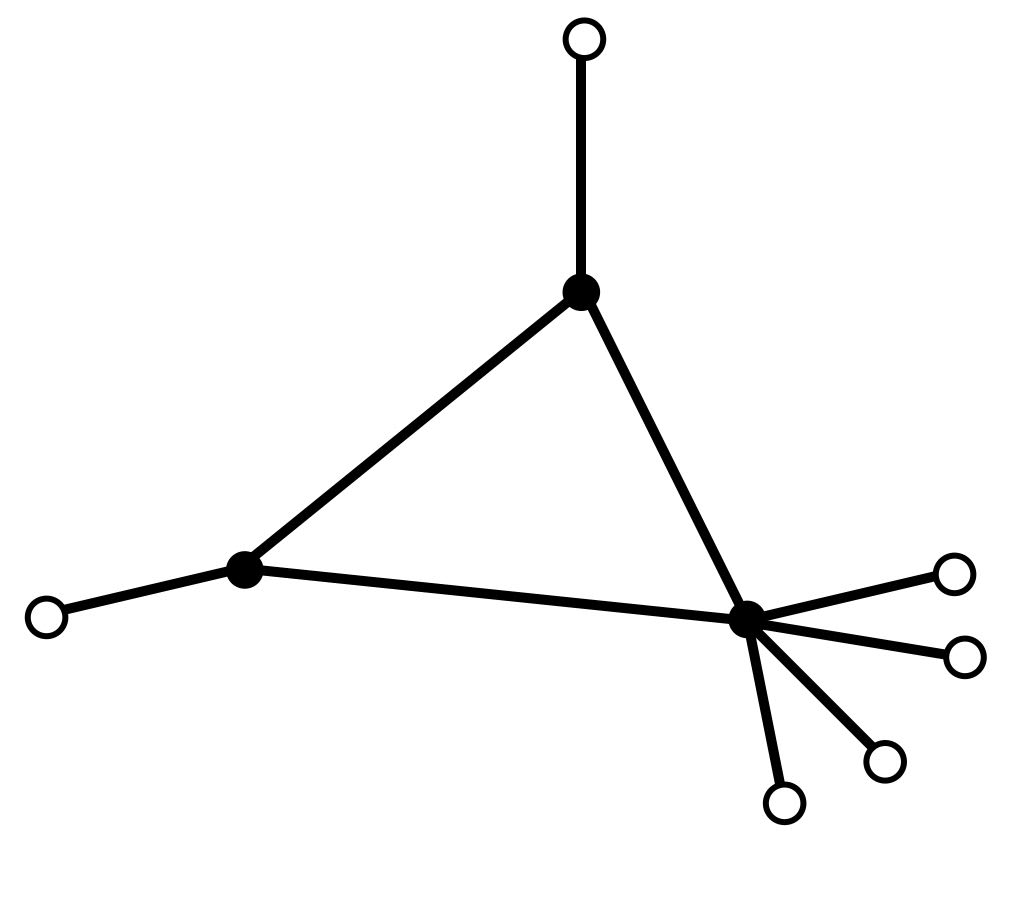}  \caption{ An example of a Feynman graph that cannot be drawn on the Loom.}\label{fig:trlegs}
\end{figure}

In fact certain graphs cannot be drawn on the Loom due to purely topological obstructions -- for example, having 'too many' legs at external vertices as discussed in \cite{Kazakov:2023nyu} -- an example of such a graph is given on figure \ref{fig:trlegs}.  Yet as we discuss in the next subsection even for these graphs Yangian symmetry can often hold, and this is usually the property we are ultimately interested in. It seems natural to expect that the combination of the two constraints \eq{con1}, \eq{con2} in fact serves as the \textit{necessary and sufficient} condition for the graph to have Yangian plus conformal symmetry, regardless of the Loom construction. This condition is also much simpler to verify or impose in practice than making reference to the Loom. It also eliminates the need to consider additional upgrades on top of the Loom construction such as `legs ending in open cells' discussed in \cite{Kazakov:2023nyu}. We believe that this simple prescription concisely describes which graphs in general (in the class of scalar planar graphs we consider) can or can not be Yangian invariant\footnote{We note that different partial answers to this general question were suggested in several different works over the recent years, including e.g.  \cite{Loebbert:2020hxk, Chicherin:2022nqq, Kazakov:2023nyu}}, incorporating known cases in a natural way. We illustrate and motivate the prescription further in several examples in the next subsection\footnote{we expect that a fully rigorous proof can be established as well but will not address this here}.

Lastly, let us note that the Loom construction can still be very useful, as it is at least guaranteed to provide Yangian invariant graphs, whereas it is not always clear for which graphs the conditions \eq{con1}, \eq{con2} are compatible (as we will also see in the next subsection).

\subsection{Examples:  Yangian symmetric graphs and 
possible obstructions}

\paragraph{Example: star/triangle graphs.} As a first example consider a conformal 3-star integral and the triangle graph, shown below:
\beq\nn
 \resizebox{0.35\hsize}{!}{
$ \vcenter{\hbox{\StarGraph}} \ , \  \vcenter{\hbox{\TriangleGraph}}$
 }
 \eeq
For the star graph, since it has no faces, the only constraint to impose is conformality, i.e. \eq{con1}. However, it is instructive to consider what happens if we rewrite this integral as a triangle by using a star-triangle transformation reviewed in appendix \ref{sec:AppendixDoubleStarIntegral}.  The resulting triangle is just a product of three propagators, but its propagator dimensions sum up to $D/2$ -- in perfect accordance with the dual constraint \eq{con2}. In fact one can verify that, starting with a triangle with general propagator powers, it is precisely when the constraint \eq{con2} is imposed that the triangle graph will satisfy an \textit{additional} differential equation in the coordinates, which can be seen to correspond to Yangian symmetry. This illustrates the role of the dual conformal constraint \eq{con2} which comes up in a somewhat unexpected way for this very simple triangle graph.

\paragraph{Example: triangle with many external legs.} As mentioned above, due to purely topological considerations not all graphs can even be drawn on the Loom, an example being the graph on figure \ref{fig:trlegs} (see \cite{Kazakov:2023nyu}). However the prescription \eq{con1}, \eq{con2} discussed above can be easily applied even to this case, and indeed one can straightforwardly verify (using e.g. the lasso method as in \cite{Kazakov:2023nyu}) that under these conditions this Feynman integral is indeed Yangian invariant.

\begin{figure}[H]
 \centering
\includegraphics[scale=0.155]{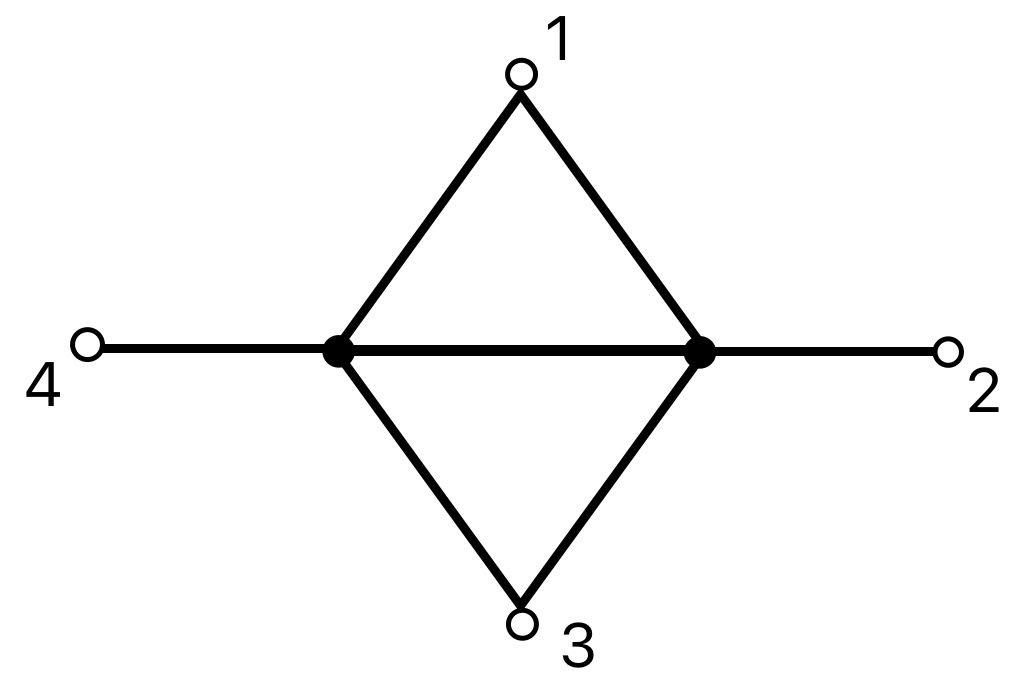}  \caption{ A ladder graph.}\label{fig:lad}
\end{figure}

\paragraph{Example: bringing together external points.} One could also ask if Yangian symmetry may allow us to constrain more involved integrals, such as the Basso-Dixon graph -- the simple $4$-point example being the ladder shown on figure \ref{fig:lad}.

To make it Yangian invariant we should impose the dual conformal symmetry constraint \eq{con2}, which in this case will apply to the two triangular faces (even though they involve an external point). Imposing it we immediately see that we can use the star-triangle transformation in such a way that  we get the 4-cross graph. The most interesting cases, like the integral in $D=4$ with all propagator powers equal to 1, lie outside this range of parameters (i.e. they do not satisfy \eq{con2}). A similar logic shows that for other ladder diagrams imposing dual conformal symmetry also leads to a drastic simplification of the graph and is not compatible with e.g. the case $D=4,\ \Delta_i=1$. In that sense, we can only directly use Yangian symmetry for the ladders when they are reducible to simpler diagrams.\footnote{ Note, however, that even for more general ladders one can still write Yangian Ward identities, which are identities of other type \cite{Corcoran:2021gda}.}.
\\

\paragraph{Example of an obstruction: Kagome graphs.} Another curious example is related to the Kagome type lattice discussed in \cite{Kazakov:2022dbd}. Here for concreteness we will consider a 6-point portion cut out of the general graph:
\begin{figure}[H]
    \centering
    \includegraphics[scale=0.1]{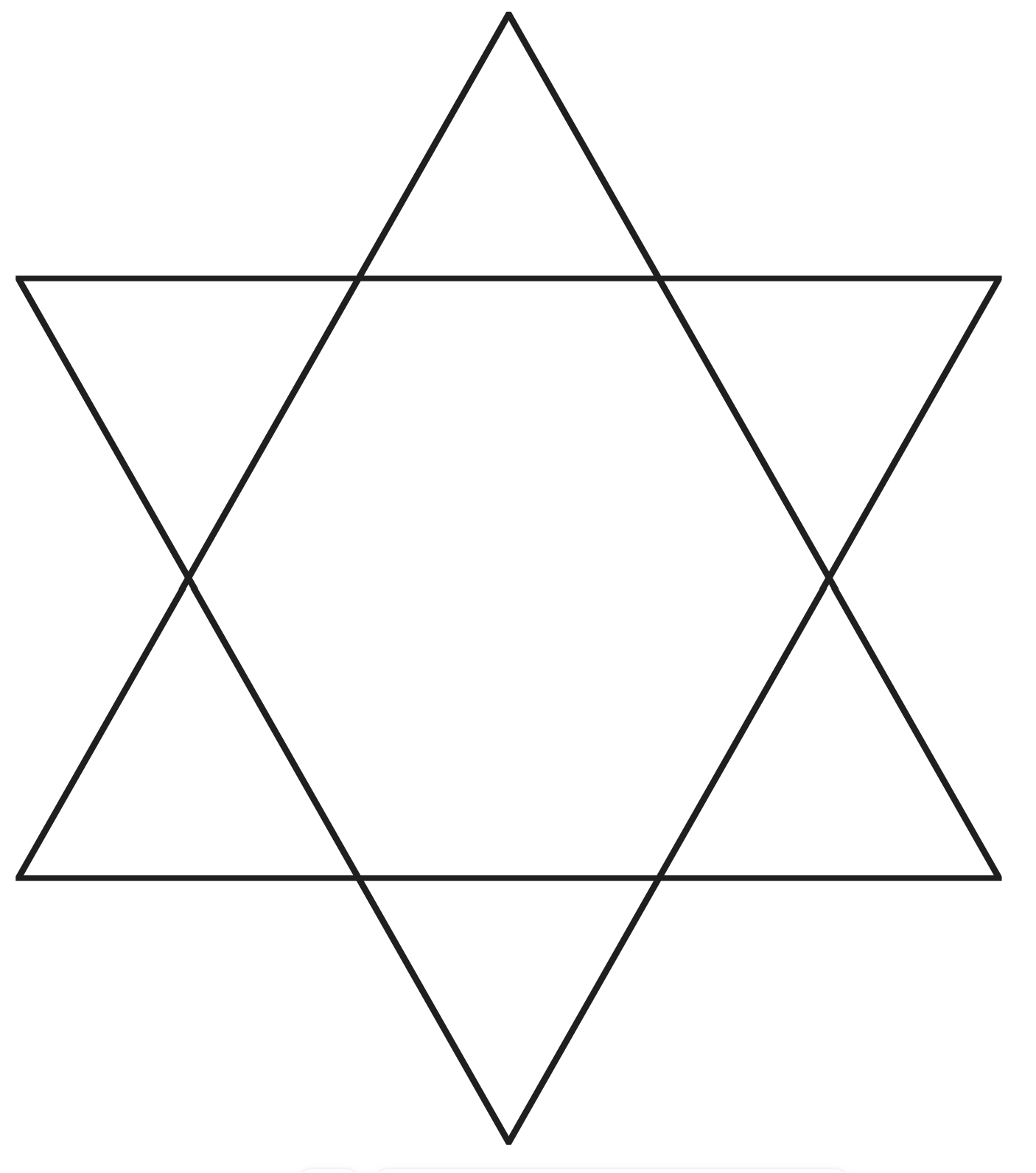}
    \caption{The 6-point star portion of the Kagome lattice graph}
    \label{fig:6ptKagome}
\end{figure}
It was observed in \cite{Kazakov:2022dbd} that if one attempts to build a Loom construction for these graphs, then one would not get a set of straight lines for the Baxter lattice. 
Let us look at this from the perspective of the general conformality and dual conformality conditions \eq{con1}, \eq{con2}. We see that there are in fact 13 conditions to be satisfied: 6 conformal conditions for vertices and 7 dual conditions for faces. There are 18 edges in this graph, however, so, naively, we have enough parameters to satisfy these constraints. Yet one can see that the constraints are not self-consistent. Indeed, first let us consider the sum of dimensions over all triangles -- it includes the sum over external edges and the edges of the hexagon:
\begin{equation}\label{eq:starfaces}
    3D = \dfrac{D}{2} \cdot 6 = \text{sum over triangles} = \text{external} + \text{hexagon} = \text{external} + 2  D 
\end{equation}
On the other hand consider the sum of conformal conditions at all the vertices -- these will include the sum over external lines and twice the sum over the edges of the hexagon:
\begin{equation}\label{eq:starvertices}
    6 \cdot D = \text{sum over vertices} =\text{external} +2 \cdot \text{hexagon} =  \text{external} +4D
\end{equation}
The two equations \eqref{eq:starfaces} and \eqref{eq:starvertices} are clearly not compatible. Hence, for this graph existence of Yangian symmetry is not possible at all for any choice of propagator powers.
\\

We see that in principle the existence of Yangian symmetry restricts the very topology of the graph. In particular, as a rule of thumb, one should think that Yangian symmetric graphs should have separated external legs. When some external legs are merged, direct Yangian symmetry is often too stringent. The conditions on dimensions in this case force it out of the physically interesting cases and make the graph reducible to simpler ones. That said, more generally  the topologies of Yangian symmetric Feynman integrals can be quite diverse and contain many important cases.

\section{Yangian equations in cross-ratio variables}

In this section we present one of our main results -- the general form of the Yangian equations in conformal cross-ratio variables. We will start by introducing the relevant notation for conformal variables, and later we will also discuss consistency conditions for the Yangian equations.

\subsection{Conformal symmetry and cross-ratios}

\label{sec:crr}

Conformal symmetry  requires that the Feynman integral should be a function of conformal invariants -- the cross ratios, up to a conformal weight factor. For $N$ points in $D$-dimensional  space, the cross ratios are defined as:
\begin{equation}
    \xi^A = \prod_{i<j} x_{ij}^{2\alpha_{ij}^A}
\end{equation}
Here
\begin{equation}
    x^2_{ij}=(x_i-x_j)^\mu (x_i-x_j)_\mu
\end{equation}
are the Poincare invariants, and $\alpha_{ij}^A$ solve the system of equations:
\begin{equation}\label{eq:SystemForAlpha}
\begin{split}
        \alpha^{A}_{ij}=\alpha^A_{ji} \, ,& \quad \alpha^A_{ii}=0
    \\
        \sum_{i=1}^N \alpha^A_{ij} = 0 \, ,& \quad \forall j =1,\ldots, N
\end{split}
\end{equation}
The index $A$ counts solutions to this system of equations, of which there are $\frac{N(N-3)}{2}$ if the dimension is large enough (as we will assume, otherwise the cross-ratios could become dependent).  In general, the number of independent cross ratios is 
\begin{equation}
    N_{cr}=\left\{ \begin{aligned}
          &    \dfrac{N(N-3)}{2} \,,  \qquad  N \leq D+2
        \\
       &ND-  \dfrac{(D+1)(D+2)}{2} \, ,  \ N \geq D+2
    \end{aligned}
    \right.
\end{equation}
In this work, we will always assume that the first of these two cases is realized.
The conformal weight prefactor is defined as:
\begin{equation}
    W_N( \mathbf{x} | \pmb{\beta} )=\prod_{i<j} x^{2\beta_{ij}}_{ij}
\end{equation}
where $\beta_{ij}$ satisfy the following conditions:
\begin{equation}\label{eq:SystemForBeta}
\begin{split}
        \beta_{ij}=\beta_{ji} \, ,& \quad \beta_{ii}=0
    \\
        \sum_{i=1}^N \beta_{ij} = - \Delta_j \, ,& \quad \forall j =1,\ldots, N
\end{split}
\end{equation}
Here $\Delta_j$ should be understood as the conformal dimensions that enter the conformal algebra operators \eq{cop}, or, in other words, the sum of powers of propagators that end at the vertex with coordinate $x_j^\mu$.

Finally, conformal symmetry implies that the Feynman integral is given as
\begin{equation}
\label{ig1}
    I_\Gamma( \mathbf{x} | \pmb{\Delta} ) = W_{N,\Gamma}(\mathbf{x}|\pmb{\beta}) I^{(0)}_\Gamma( \pmb{\xi} | \pmb{\Delta} )
\end{equation}
The function in the r.h.s. depends only on $N_{cr}$ independent cross ratios, corresponding to some choice of basis in the solution space to \eqref{eq:SystemForAlpha}. The explicit form of the function will of course depend on this choice. 
Note that there is an ambiguity in the choice of $\beta_{ij}$ -- one can always add a solution to the homogenous equations to any solution of \eqref{eq:SystemForBeta}. This, however, is compensated with a redefinition of the $I^{(0)}$ function by multiplying it with the corresponding product of cross ratios.

\subsection{Derivation of the Yangian equations in cross-ratios}

Given the form of the general conformally invariant Feynman diagram in the previous section, it is natural to ask how the Yangian symmetry acts when written in conformal cross-ratio variables. In this subsection we derive an explicit expression for this action.

First, let us write the level-one momentum generator explicitly as a differential operator:
\begin{equation}
\begin{split}
    (-i)\widehat{P}^\mu  = \dfrac{1}{2}\sum_{i<j}&\left( \delta^{\mu \alpha}\delta^{\lambda \nu} - \delta^{\nu \alpha} \delta^{\mu\lambda} - \delta^{\mu \nu} \delta^{\alpha\lambda}  \right) (x_j-x_i)^{\alpha} \dfrac{\partial^2}{\partial x_i^\lambda \partial x_j^\nu} + \dfrac{1}{2} \sum_{i<j}  \left(  \Delta_i \dfrac{\partial }{\partial x^{\mu}_j} -  \Delta_j \dfrac{\partial }{\partial x^{\mu}_i}  \right)  - \\
& -\sum_{i} s_i \dfrac{\partial}{\partial x_i^\mu}. \label{ipmu}  
\end{split}
\end{equation}
Throughout the discussion in this subsection we take the evaluation parameters  $s_i$ to be arbitrary.
It was previously proposed in \cite{Loebbert:2019vcj}, based on a variety of examples, that one has in general\footnote{in our notation the operators $\PDE_{ik}$ differ by an overall constant numerical factor from those in \cite{Loebbert:2019vcj}}
\begin{equation}\label{eq:YangianToPDE}
     (-i){\widehat P}^\mu \, I_\Gamma(\mathbf{x}|\pmb{\Delta}) = W_{N,\Gamma}(\mathbf{x}|\pmb{\beta}) \sum_{i<j} \dfrac{x_{ik}^\mu}{x_{ik}^2} \PDE_{ik} I^{(0)}_\Gamma( \pmb{\xi} | \pmb{\Delta} )
\end{equation}
Here the differential operators $\PDE_{ik}$ are written only in terms of cross-ratios. If the dimension $D$ is large enough, then all the vectors $\dfrac{x_{ik}^\mu}{x_{ik}^2}$ are independent, and we conclude that Yangian symmetry implies
\begin{equation}\label{eq:PDE}
    \PDE_{ik} I^{(0)}_\Gamma( \pmb{\xi} | \pmb{\Delta} ) = 0 \,.
\end{equation}
As one may notice straight away, the system has more equations than there are variables. This means that not all the equations should be independent. Indeed, for the studied examples, with specific graphs and, hence, fixed $s_i$ one can observe, that the number of independent equations is always lower and coincides with the number of $\xi^A$ variables. In fact the equations turn out to have a finite dimensional space of solutions, i.e. they fix the value of the integral up to a finite number of integration constants, that have to be determined by symmetries and boundary conditions. While examples of equations for a few graphs where studied, the general form was unknown.
\\

Below we present a derivation of the general form for the operators $\PDE_{ik}$, proving the structure \eqref{eq:YangianToPDE} along the way. The results of expression are independent of the specific choice of cross-ration, i.e. the choice of $\alpha_{ij}^A$, and are written in a way that is covariant. In short, the main idea is to go through an intermediate stage in the derivation. If we impose only the Poincare symmetry, we get that the Feynman integral depends on the lengths $x_{ij}^2$. So, first we rewrite the level-one generator in these variables. Only then we impose the conformal symmetry which further restricts that the variables $x_{ij}^2$ should only enter as cross ratios.

Let us now sketch the derivation itself. First, we notice that while the derivatives in \eq{ipmu} act on individual coordinates, the Feynman graph is a function only of their squared differences $x_{ij}^2$. We will assume that the number of points is low enough compared to the dimension, so we can view $x_{ij}^2$ (with $i<j$) as independent variables. Thus we can rewrite the derivatives to act on these variables, for example we have
\beq
    \frac{\d}{\d x_i^\mu}=2\sum_{k\neq i}x_{ik}^\mu\frac{\d}{\d x_{ik}^2}
\eeq
Proceeding in this way, we find that when acting on the Feynman graph, \eq{ipmu} takes the form
\beq
    (-i){\widehat P}^\mu =\sum_ix_{1i}^\mu \cdot{\rm{PDE}}_i(x^2)
\eeq
where we indicate that the operators ${\rm{PDE}_i}$ act on the variables $x_{kl}^2$. Explicitly, we find that they have the form 
\begin{equation}
\begin{split}
        {\mathrm{PDE}}_i &= \sum_{\substack{j>i \\
    k \neq i ,
    l \neq j}} (x_{jk}^2 +x_{jl}^2 - x_{kl}^2 ) \dfrac{\partial^2}{\partial (x_{ik}^2 ) \partial (x^{2}_{jl})} + \dots 
\end{split}
\end{equation}
where the dots indicate several similar terms,  as well as simpler terms with only the first derivatives $\frac{\d}{\d x_{ij}^2}$. We do not write these terms here for the sake of readability but they are straightforward, if laborious, to compute.

Next, we recall that the Feynman graph on which we act has the structure \eq{ig1}, i.e. a prefactor times a function of conformal cross-ratios. When acting on this with e.g. $\frac{\d}{\d x_{ij}^2}$ we find that the derivatives can hit either the prefactor or the nontrivial remaining function, and the result can be schematically written as 
\beq
    \frac{\d}{\d x_{ij}^2}\to \theta_{ij}
\eeq
where $\theta_{ij}$ are the differential operators in cross-ratios defined as
\beq
    \theta_{ij}=\sum_A\alpha_{ij}^A\xi^A\frac{\d}{\d\xi^A}+\beta_{ij}
\eeq
When acting with the 2nd derivatives $\dfrac{\partial^2}{\partial (x_{ik}^2 ) \partial (x^{2}_{jl})}$ there will also be nontrivial cross-terms. Combining everything together, we see that we mainly get sums of terms of the kind
\begin{equation}
    T_{ijkl}=(x_{jk}^2 +x_{jl}^2 - x_{kl}^2 ) \dfrac{1}{x_{ik}^2 x^2_{jl}} \theta_{ik}\theta_{jl}  
\end{equation}
which we can rewrite in terms of cross-ratios as
\beq
     T_{ijkl}= \left( \dfrac{\chi_{ilkj} }{x^2_{il}} + \dfrac{1}{x_{ik}^2} - \dfrac{\chi_{ijkl}}{x^2_{ij} }  \right)\theta_{ik}\theta_{jl}
\eeq
Here $\chi_{ijkl}$ are the standard 4-point cross-ratios,
\begin{equation}
\chi_{ijkl}=\dfrac{x^2_{ij}x^2_{kl}}{x^2_{ik}x^2_{jl}}
\end{equation}
Note that the $\chi$'s of course are expressible in terms of our basis cross-ratios $\xi^A$, with the exact expression depending on the chosen basis.

Combining all these steps, we find, after rather lengthy calculations, that when acting on our graph we have
\beq
    (-i){\widehat P}^\mu =\sum_ix_{1i}^\mu\sum_{k\neq i}\frac{{\rm{PDE}}_{ik}(\xi)}{x_{ik}^2} 
\eeq
where the differential operators ${\rm{PDE}}_{ik}$ act now only on the conformal cross-ratios $\xi^A$ (and not on the prefactor of the Feynman graph). This can be further rearranged by observing that they are antisymmetric, 
${\rm{PDE}}_{ik}=-{\rm{PDE}}_{ki}$, and the trivial identity
$x_{1i}^\mu=x_{1k}^\mu+x_{ki}^\mu$. We find
\beq
\label{xik2}
    (-i){\widehat P}^\mu =-2\sum_{i<k}\frac{x_{ik}^\mu}{x_{ik}^2}{\rm{PDE}}_{ik}(\xi)
\eeq
Finally, by the argument similar to that in \cite{Loebbert:2019vcj}  we expect that at least for large enough dimension the vectors $\frac{x_{ik}^\mu}{x_{ik}^2}$ should be independent, and therefore their coefficients in \eq{xik2} have to vanish, i.e. the operators ${\rm{PDE}}_{ik}$ annihilate our graph. The explicit form of these operators is the main outcome of our long calculation. It is given in \eq{eq:PDEik} in the Introduction, and we repeat it here:
\begin{equation}\label{eq:PDEik2}
\begin{split}
        \PDE_{ik} &= 2\left( \sum_{l>j>i} - \sum_{l<j<i} + \sum_{l<k<i ; j} - \sum_{l>k>i; j } \right) \chi_{iklj} \theta_{il}\theta_{jk} +\sum_{j\neq i}(\delta_{j>i}-\delta_{j<i})\theta_{ik}\theta_{ij} 
        \\
&\;+\left(\delta_{i<k}\left(\Delta_k-D \right)- \delta_{i>k}\left(\Delta_i-D \right)  \right)\theta_{ik} + 2(s_i-s_k)\theta_{ik}
\end{split}
\end{equation}
This constitutes one of our main results. 

We have verified that  equations \eq{eq:PDEik2} perfectly match known data in the literature:

\begin{itemize}
\item For the 4pt cross integral in any $D$ with generic scaling dimensions $\Delta_k$, $k=1,\dots,4$ (summing up to $D$) we reproduce the equations from \cite{Loebbert:2019vcj}.
    \item For the double cross integral with six external points in $D=4$ with all scaling dimensions $\Delta_k=1$ we reproduce equations (79) from \cite{Loebbert:2019vcj}.
    \item For the 6pt cross in any $D$ with any $\Delta$'s (again summing up to $D$) we reproduce the highly involved equations (E1)-(E15) in \cite{Loebbert:2019vcj}.
\end{itemize} 
These rather nontrivial and constraining checks serve as an important verification of our key results.

Let us finally mention again that strictly speaking the equations we derive are valid for 'large enough' dimension, which guarantees independence of the relevant vectors we used at several stages. It would be interesting to investigate in the future what happens if this assumption does not hold, using the equations we derived as an important starting point from which to explore some reduction.

\subsection{Consistency and non-reducibility  conditions}
\label{sec:cons}

So far in this section we have worked with a generic choice of the evaluation parameters $s_i$ and the scaling dimensions $\Delta_i$ of the external legs. Yet, as these parameters enter the Yangian equations explicitly, it is natural  to expect that consistency of the equations could lead to constraints for their values (in particular since we typically have more equations than independent variables, i.e. cross-ratios). In this subsection we derive these constraints explicitly for the 4-pt case, and also discuss what happens more generally. These constraints will play an important role in relating the Yangian to GKZ systems in section 5.

In addition to consistency of the equations, we will also impose the property we call non-reducibility -- namely, we require that our Feynman integrals do not satisfy 1st order equations in the cross-ratios. This excludes cases which are in a sense `trivial' -- e.g. disconnected graphs or simple products of propagators connecting the external points. It seems to leave only the truly nontrivial and substantive cases, described typically by hypergeometric functions, and will be important for the general statements we make in section 5 below. From now on, in this and the next sections we will consider only non-reducible graphs. Thus, in total we impose two types of constraints: consistency of the equations and non-reducibility of the graph.

Let us see what these requirements give for the simplest case of 4pt integrals. We start with an unspecified Feynman integral that we assume to be conformal and Yangian invariant, keeping generic $s_i$ and $\Delta_i$. It can be written as 
\beq 
\label{g4f}
I(x_1,x_2,x_3,x_4)=x_{24}^{-2\Delta_2}x_{14}^{\Delta_3-\Delta_4-\Delta_1+\Delta_2}x_{13}^{-\Delta_3+\Delta_4-\Delta_1-\Delta_2}x_{34}^{-\Delta_3-\Delta_4+\Delta_1+\Delta_2}\phi(u,v)
\eeq
where we take the cross-ratios to be
\beq
\label{uv1}
u=\frac{x_{12}^2x_{34}^2}{x_{13}^2x_{24}^2} \ , \ \ \ \ v=\frac{x_{14}^2x_{23}^2}{x_{13}^2x_{24}^2}
\eeq
Then our general equations \eq{eq:PDEik2} can be straightforwardly written explicitly, and they give six nontrivial 2nd order differential equations for the function $\phi(u,v)$. By making their linear combinations we can eliminate some of the derivatives and potentially find 1st order equations for the function $\phi(u,v)$. We should prohibit the existence of these 1st order equations (i.e. we require that their coefficients vanish), since  we impose non-reducibility as mentioned above. As a result we find that $s_i$ are uniquely fixed, and furthermore the $\Delta$'s must satisfy one of the three conditions:
\beq
\label{4ptc}
\Delta_1+\Delta_2+\Delta_3+\Delta_4=D \ \ \text{or} \ \ 
    \Delta_1+\Delta_2=\Delta_3+\Delta_4 \ \ \text{or} \ \ \Delta_1+\Delta_4=\Delta_2+\Delta_3
\eeq

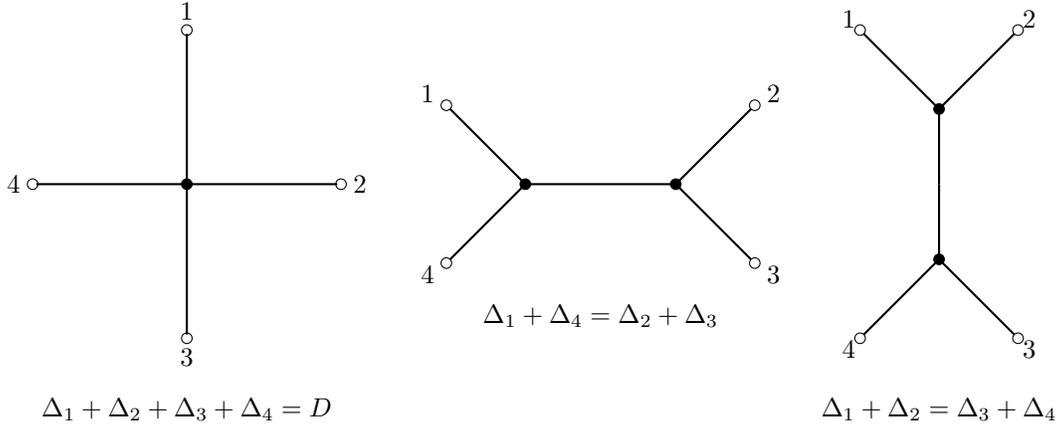
\begin{figure}[H]
 \centering

\begin{tikzpicture}
    \begin{scope}[shift={(-5,0)}] 
        \draw[thick] (0,0) -- (0,2); 
        \draw[thick] (0,0) -- (0,-2); 
        \draw[thick] (0,0) -- (2,0); 
        \draw[thick] (0,0) -- (-2,0); 
        \filldraw (0,0) circle (2pt); 
        \draw (0,2.05) circle (2pt); 
        \draw (2.05,0) circle (2pt); 
        \draw (0,-2.05) circle (2pt); 
        \draw (-2.05,0) circle (2pt); 
        \node at (0,2.3) {$1$}; 
        \node at (2.3,0) {$2$}; 
        \node at (0,-2.3) {$3$}; 
        \node at (-2.3,0) {$4$}; 
        \node at (0,-3) {$\Delta_1 + \Delta_2 + \Delta_3 + \Delta_4 = D$};
    \end{scope}

    \begin{scope}[shift={(0.5,0)}] 
        \draw[thick] (-1,0) -- (0,0) -- (1,0); 
        \draw[thick] (-1,0) -- (-2,1); 
        \draw[thick] (-1,0) -- (-2,-1); 
        \draw[thick] (1,0) -- (2,1); 
        \draw[thick] (1,0) -- (2,-1); 
        \filldraw (-1,0) circle (2pt); 
        \filldraw (1,0) circle (2pt); 
        \draw (-2.05,1.05) circle (2pt); 
        \draw (-2.05,-1.05) circle (2pt); 
        \draw (2.05,1.05) circle (2pt); 
        \draw (2.05,-1.05) circle (2pt); 
        \node at (-2.3,1.2) {$1$};
        \node at (2.3,1.2) {$2$};
        \node at (-2.3,-1.2) {$4$};
        \node at (2.3,-1.2) {$3$};
        \node at (0,-1.75) {$\Delta_1 + \Delta_4 = \Delta_2 + \Delta_3$};
    \end{scope}

    \begin{scope}[shift={(5,0)}] 
    \draw[thick] (0,0) -- (0,1); 
    \draw[thick] (0,0) -- (0,-1); 
    
    \draw[thick] (0,1) -- (-1,2); 
    \draw[thick] (0,1) -- (1,2); 
    
    \draw[thick] (0,-1) -- (-1,-2); 
    \draw[thick] (0,-1) -- (1,-2); 
    
    \filldraw (0,1) circle (2pt); 
    \filldraw (0,-1) circle (2pt); 
    \draw (-1.05,2.05) circle (2pt); 
    \draw (1.05,2.05) circle (2pt); 
    \draw (-1.05,-2.05) circle (2pt); 
    \draw (1.05,-2.05) circle (2pt); 
    
    \node at (-1.2,2.2) {$1$}; 
    \node at (1.2,2.2) {$2$}; 
    \node at (-1.2,-2.2) {$4$}; 
    \node at (1.2,-2.2) {$3$}; 
    
    \node at (0,-3) {$\Delta_1 + \Delta_2 = \Delta_3 + \Delta_4$}; 
    \end{scope}
\end{tikzpicture}
\caption{The standard non-reducible 4-point graphs.}\label{fig:4pt2}
\end{figure}

These conditions are precisely the ones realized for the simplest 4pt Yangian invariant graphs -- respectively, the cross integral and two versions of the 4pt double-cross integral, all of which are shown on figure \ref{fig:4pt2}. Our analysis means that they in fact must be satisfied for \textit{any} Yangian invariant non-reducible 4pt integral. 

\subsubsection{Examples of reducible 4pt graphs}

It may appear somewhat surprising that any Yangian invariant 4pt integral must satisfy the constraints \eq{4ptc} unless it essentially trivializes (is reducible in our terminology). 
In order to better understand these constraints let us give a few examples of graphs that violate them, and show that they are indeed reducible. 

\bigskip

\paragraph{Example 1 -- simple square.} The first example we consider is the `square' graph, given by  four propagators:
\begin{center}
   \begin{tikzpicture}
   \foreach \i in {1,...,4}
    \draw[thick] (360*\i/4:0.75cm) --(360*\i/4+360/4:0.75cm);
    \foreach \i in {1,...,4}
    \draw[fill=white,thick] (360*\i/4:0.75cm) circle(1.5pt);
    \foreach \i in {1,...,4}
    \node (\i) at (-360*\i/4:1.05cm) {\i};
\end{tikzpicture} 
\end{center}
The corresponding Feynman integral is not even an integral, but simply a product of propagators:
\begin{equation}
    I_{\scalebox{0.15}{\SquareGraph} } = \dfrac{1}{x^{2\Delta_{12}}_{12}x^{2\Delta_{23}}_{23} x^{2\Delta_{34}}_{34} x^{2\Delta_{14}}_{14} }
\end{equation}
One can straightforwardly check that this graph is Yangian invariant if
\begin{equation}
    \sum_{i=1}^4 \Delta_{i}  = 2D
\end{equation}
which is clearly \textit{not} one of the constraints we found above in \eq{4ptc}. However, there is no contradiction as this case violates our assumption about non-reducibility, i.e. absence of first order equations satisfied by the integral. The corresponding function of the conformal cross-ratios is simply a constant:
\begin{equation}
    I^{(0)}_{\scalebox{0.15}{\SquareGraph} }(u,v | \pmb{\Delta})  = 1
\end{equation}
and obviously satisfies 1st order equations such as $\d_u I=0, \ \d_vI=0$.

\paragraph{Example 2 -- square with external legs.} Another instructive case is the square with four external legs:
\begin{center}
   \begin{tikzpicture}
   \foreach \i in {1,...,4}
    \draw[thick] (360*\i/4+360/8:0.75cm) --(360*\i/4+360/4+360/8:0.75cm);
    \foreach \i in {1,...,4}
    \draw[thick,externalline] (-360*\i/4+360/8:0.75cm) --+ (-360*\i/4+360/8:0.75cm) node[pos=1.4] {\i};
\end{tikzpicture} 
\end{center}
 If we try to draw this graph on the Loom (or in another way) we find that Yangian symmetry is only possible if the integral propagator dimensions sum up to $D$. Together with conformal symmetry this gives two conditions for external dimensions:
\begin{equation}
\label{sql1}
    \sum_{i=1}^4 \Delta_{i}  = 2D\,, \quad \Delta_1+\Delta_3=\Delta_2+\Delta_4
\end{equation}
These are both not among our constraints \eqref{4ptc}.
At the same time, unlike the previous example, at first sight this graph corresponds to a non-trivial integral, as it has four integration vertices. Let us see that in fact it can be trivialized by using star-triangle transformations (reviewed in appendix \ref{sec:AppendixDoubleStarIntegral}). Applying it to the ''stars'' connected to external vertices $2$ and $4$, and then to the remaining two vertices, we get the following sequence of transformations:
\begin{center}
   \begin{tikzpicture}[>={Stealth}]
   \begin{scope}
       \foreach \i in {1,...,4}
    \draw[thick] (360*\i/4+360/8:0.75cm) --(360*\i/4+360/4+360/8:0.75cm);
    \foreach \i in {1,...,4}
    \draw[thick,externalline] (-360*\i/4+5*360/8:0.75cm) --+ (-360*\i/4+5*360/8:0.75cm) node[pos=1.4] {\i};
   \end{scope}

   \begin{scope}[xshift=4.5cm]
           \draw[thick,externalline] (360/4+360/8:0.75cm) --+ (360/4+360/8:0.75cm) node[pos=1.4] {1};
            \path (360/8:0.75cm) --+ (360/8:0.75cm) node[pos=1.4] {2};
        \draw[thick,externalline] (-360/8:0.75cm) --+ (-360/8:0.75cm) node[pos=1.4] {3};
        \path (-360/4-360/8:0.75cm) --+ (-360/4-360/8:0.75cm) node[pos=1.4] {4};
        \draw[thick] (360/4+360/8:0.75cm) -- (360/8:1.5cm);
        \draw[thick] (-360/8:0.75cm) -- (360/8:1.5cm);
        \draw[thick] (360/4+360/8:0.75cm) -- (360/8-360/2:1.5cm);
        \draw[thick] (-360/8:0.75cm) -- (360/8-360/2:1.5cm);
        \draw[thick] (-360/8:0.75cm) to [bend left =12.5] (360/8+360/4:0.75cm);
        \draw[thick] (-360/8:0.75cm)  to [bend right =12.5] (360/8+360/4:0.75cm);
   \end{scope}
     \begin{scope}[xshift=9cm]
           \draw[thick,externalline] (360/4+360/8:0.75cm) --+ (360/4+360/8:0.75cm) node[pos=1.4] {1};
            \path (360/8:0.75cm) --+ (360/8:0.75cm) node[pos=1.4] {2};
        \draw[thick,externalline] (-360/8:0.75cm) --+ (-360/8:0.75cm) node[pos=1.4] {3};
        \path (-360/4-360/8:0.75cm) --+ (-360/4-360/8:0.75cm) node[pos=1.4] {4};
        \draw[thick] (360/4+360/8:0.75cm) -- (360/8:1.5cm);
        \draw[thick] (-360/8:0.75cm) -- (360/8:1.5cm);
        \draw[thick] (360/4+360/8:0.75cm) -- (360/8-360/2:1.5cm);
        \draw[thick] (-360/8:0.75cm) -- (360/8-360/2:1.5cm);
   \end{scope}
     \begin{scope}[xshift=2.25cm+4.5cm,yshift=-4.5cm]
           \foreach \i in {1,...,4}
    \draw[thick] (360*\i/4+360/8:0.75cm) --(360*\i/4+360/4+360/8:0.75cm);
    \foreach \i in {1,...,4}
    \draw[fill=white,thick] (360*\i/4+360/8:0.75cm) circle(1.15pt);
    \foreach \i in {1,...,4}
    \node (\i) at (-360*\i/4+5*360/8:1.05cm) {\i};
    \draw[thick] (360/8-180:0.75cm) to [bend left =10] (360/8:0.75cm);
    \draw[thick] (360/8-180:0.75cm) to [bend right =10] (360/8:0.75cm);
   \end{scope}
    \begin{scope}[xshift=2.25cm,yshift=-4.5cm]
           \foreach \i in {1,...,4}
    \draw[thick] (360*\i/4+360/8:0.75cm) --(360*\i/4+360/4+360/8:0.75cm);
    \foreach \i in {1,...,4}
    \draw[fill=white,thick] (360*\i/4+360/8:0.75cm) circle(1.15pt);
    \foreach \i in {1,...,4}
    \node (\i) at (-360*\i/4+5*360/8:1.05cm) {\i};
   \end{scope}
   \draw[thick,->] (2cm,0) to (4.5cm-2cm,0);
   \draw[thick,->] (2cm+4.5cm,0) to (4.5cm-2cm+4.5cm,0);
   \draw[thick,<-] (2cm+2.25cm,-4.5cm) to (4.5cm-2cm+2.25cm,-4.5cm);
    \draw[thick,->,shorten >=0.75cm,shorten <=0.25cm] (2.5cm+4.5cm+2.25cm,-2.25cm) to [bend left=25] (3.6cm+4.5cm,-4.5cm);
   
\end{tikzpicture} 
\end{center}

The second and the last steps in this sequence include collapsing two propagators. At this stage the constraints \eq{sql1} on dimensions are crucial -- they imply that the sum of the dimensions for these two propagators is zero and indeed the propagators can be removed. Otherwise, one could not simplify the graph further after the first step. We see that with a series of star-triangle transformation we have reduced the graph to the simple square graph. This means that our graph again violates the non-reducibility assumption so that again there is no contradiction. Explicitly, keeping track of all the transformations, we find that up to constant prefactors this integral results in
\begin{equation}
   I_{\scalebox{0.15}{\SquareWithExternalLinesGraph} }(\pmb{x} | \pmb{\Delta})  \sim \dfrac{1}{x_{12}^{2(\Delta_2-\Delta_3)} x_{23}^{2\Delta_3} x_{14}^{2\Delta_4} } \left( \dfrac{x_{12}^2 x_{34}^2}{x_{14}^2 x_{23}^2} \right)^a 
\end{equation}
and hence
\begin{equation}
    I^{(0)}_{\scalebox{0.15}{\SquareWithExternalLinesGraph} }(\pmb{\xi} | \pmb{\Delta})=\left( \xi^{1} \right)^a \,
\end{equation}
where $a$ is some combination involving the internal lines' dimensions and we have chosen the cross ratios as
\begin{equation}
    \xi^1 =\dfrac{x_{12}^2 x_{34}^2}{x_{14}^2 x_{23}^2}  \, ,\quad \xi^2= \dfrac{x_{13}^2 x_{24}^2}{x_{14}^2 x_{23}^2} \,.
\end{equation}
Thus we see that this example again violates the assumption about non-vanishing first derivatives since we have
\begin{equation}
    \dfrac{\partial}{\partial \xi^2}  I^{(0)}_{\scalebox{0.15}{\SquareWithExternalLinesGraph} }(\pmb{\xi} | \pmb{\Delta})  = 0  \, . 
\end{equation}
At the same time we again observe that this violation leads to a kind of trivialization of the integral, where it is reduced to a much simpler one by star-triangle transformations

\subsubsection{Comments on higher-point cases}

In principle one could try to repeat the same analysis for higher-point Yangian equations. However even for 5 points this becomes rather difficult technically and so far we were able to only get partial results. Instead, in the next section we propose as a conjecture an alternative way that gives a shortcut to derive the consistency and non-reducibility conditions, which we have checked against our partial results for the 5pt and 6pt cases.

\section{Relation to GKZ equations}

In this section we will demonstrate that under certain conditions the system of Yangian equations is equivalent to a very specific GKZ system. 
\\

First notice that the second order part of the differential operators in the previous section is essentially a sum of terms that have the form:
\begin{equation}
\label{lh}
    \hat{\mathcal{L}}_{iklj}= \theta_{ik}\theta_{lj} - \chi_{iklj}\theta_{il}\theta_{kj}
\end{equation}
These differential operators actually appear as a result of the reduction to conformal variables of simple operators in Poincare invariant variables:
\begin{equation}\label{eq:LLoperators}
    \LL_{iklj}=\dfrac{\partial^2}{\partial x^2_{ik} \partial x^2_{lj} }- \dfrac{\partial^2}{\partial x^2_{il}  \partial x^2_{kj}}
\end{equation}
I.e.
\begin{equation}
    \LL_{iklj}   I_\Gamma( \mathbf{x} | \pmb{\Delta} ) =\dfrac{ W_{N,\Gamma} (\mathbf{x}|\pmb{\beta}) }{x^2_{ik} x^2_{lj}} \LLL_{iklj} I^{(0)}_\Gamma( \pmb{\xi} )
\end{equation}
We will explain below how these operators $\hat L_{iklj}$ actually belong to a certain GKZ ideal and conjecturally generate it as well. Before that we study the exact relation between the system of equations generated by $\PDE_{ik}$ and the system of operators $ \LL_{iklj}$. 

\subsection{Recasting the Yangian equations in GKZ terms}

\label{sec:ycons}

Let us see how we can make explicit a GKZ-like part in the Yangian equations. In order to 
 observe more structures, let us define a generalized version of the GKZ-type operators \eq{lh} where we extract or add some propagators from the graph before applying the differential operator (and undo this operation afterward). In particular, this is motivated by 4pt graphs like those on figure \ref{fig:4pt2} which reduce to the cross integral (and thus satisfy the corresponding GKZ system for the cross discussed e.g. in \cite{Pal:2023kgu}), but only after we apply the star-triangle relation to one of the vertices, which creates extra propagators that need to be removed, see Appendix \ref{sec:AppendixDoubleStarIntegral} for details. Thus, we consider the operators
\begin{equation}
\label{lk}
    L^{\kappa}_{iklj}= \left(\prod_{m<n}x_{mn}^{2\kappa_{mn} } \right)^{-1}   L_{iklj} \left(\prod_{m<n}x_{mn}^{2\kappa_{mn} } \right)
\end{equation}
and their counterpart $\mathcal{L}^{\kappa}_{iklj} $, where:
\begin{equation}\label{eq:LLconjugate}
    \mathcal{L}^{\kappa}_{iklj}  = \theta^{\kappa}_{ik}\theta^{\kappa}_{lj} - \chi_{iklj}\theta^{\kappa}_{il}\theta^{\kappa}_{kj} \, , \qquad  \theta^{\kappa}_{ij} = \theta_{ij}+ \kappa_{ij}
\end{equation}
Using the explicit form of Yangian equations ${\rm PDE}_{ik}$ derived above, we find that they can be recast in terms of these operators up to extra terms containing only 1st order derivatives:
\begin{equation}
\label{lr}
\mathrm{PDE}_{ik} = {\cal L}_{ik}^\kappa + R_{ik}^\kappa
\end{equation}
where the first part contains our GKZ-like operators, 
\beq
    {\cal L}_{ik}^\kappa = 2\left( \sum_{l>j>i} - \sum_{l<j<i} + \sum_{l<k<i , j} - \sum_{l>k>i, j } \right) \hat{\mathcal{L}}^{\kappa}_{iklj} 
\eeq
and the remaining second part $R_{ik}^\kappa$ contains only first derivatives in cross-ratios\footnote{and also terms without derivatives}. Already this is a nontrivial observation which recasts the most complicated, in a sense, part of the Yangian equations in terms of GKZ operators up to a simple conjugation in \eq{lk}.

Next, suppose the remainder $R_{ik}^\kappa$ vanishes. Then since the Yangian operators annihilate the graph, the same must be true for the operators ${\cal L}_{ik}^\kappa$. Furthermore we observe that at least up to 5 points the set of independent individual operators $\hat L_{ijkl}$ can be linearly expressed in terms of $L_{ik}^\kappa$ and thus the GKZ-type operators $\hat L_{ijkl}$ also annihilate the Feynman integral. So, up to a simple conjugation by propagator factors, the Feynman integral in these cases satisfies the GKZ equations.\footnote{  The functions in the kernels of the conjugated \eqref{eq:LLconjugate} and the original operators \eqref{eq:LLoperators} clearly only differ by the $\kappa$-dependent prefactor and thus are trivially mapped to each other.} For higher number of points the reduction to individual $\hat L_{ijkl}$ operators will not be as simple, but taking into account the functional dependencies between various operators we expect it will be true at least for a large class of graphs. Let us sketch a possible argument to support this. As we will later see, the number of independent differential operators $\hat L_{ijkl}$ (the dimension of the respective differential ideal) is equal to the number of cross ratios $N_{cr}=N(N-3)/2$. This means that the operator system of equations $\PDE_{ik}={\cal L}_{ik}^\kappa$ is overdetermined, as there are naively $\frac{N(N-1)}{2}$ equations. However, if we assume that the system $\PDE_{ik}$ is not overdetermined, then there would be also only $N_{cr}$ independent operators. Hence, the system would allow expressing the GKZ operators from the $\PDE$'s (and vice versa), which would mean that individual  $\hat L_{ijkl}$  operators are also annihilators of the Yangian invariant integral. Making this logic precise and uncovering the details requires a detailed exploration, which we leave for the future.

\bigskip

We see that in the cases when the remainder $R_{ik}^\kappa$ in \eq{lk} vanishes, the Yangian is rewritten completely in GKZ terms. What are these cases precisely? We have found a number of situations when this happens which we list below.

\paragraph{Case 1: four external points.} For four external points, we find that the remainder vanishes if and only if the consistency conditions from \eq{4ptc} are satisfied. The parameters $s_i$ and $\kappa$ are then also fixed.  In other words, vanishing of the remainder is equivalent to the graph being non-trivial (in the sense explained in section \ref{sec:cons}). 

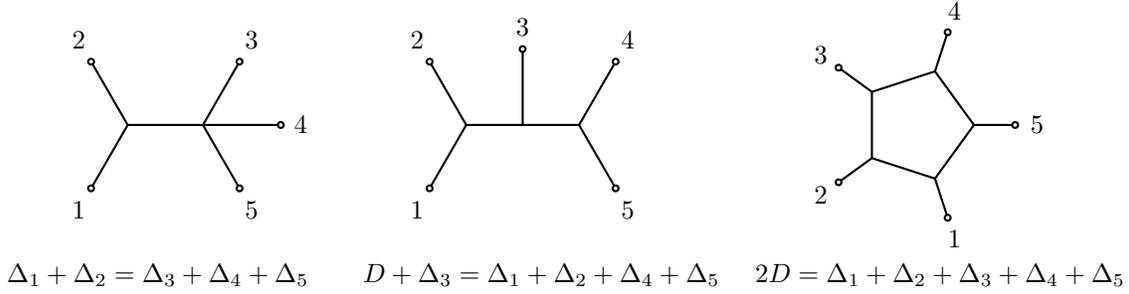
\begin{figure}[H]
\begin{center}
       \begin{tikzpicture}
\begin{scope}
    \node (1) at ($(-1cm,0)+(-90-30:1.3cm)$) {$1$};
     \node (2) at ($(-1cm,0)+(90+30:1.3cm)$) {$2$};
   \node (3) at (60:1.3cm) {$3$};
     \node (4) at (0:1.3cm) {$4$};
    \node (5) at (-60:1.3cm) {$5$};
    
     \draw[thick,externalline] (-1cm,0) -- (1);
      \draw[thick,externalline] (-1cm,0) -- (2);
    \draw[thick,externalline] (0,0) -- (3);
    \draw[thick,externalline] (0,0) -- (4);
     \draw[thick,externalline] (0,0) -- (5);
     \draw[thick] (0,0)-- +(-1cm,0);   \node (cond) at (-0.6,-2) {$\Delta_1+\Delta_2=\Delta_3+\Delta_4+\Delta_5$} ;   
\end{scope}
    
     \begin{scope}[xshift=5cm]
         \node (1) at ($(-1.5cm,0)+(-90-30:1.3cm)$) {$1$};
     \node (2) at ($(-1.5cm,0)+(90+30:1.3cm)$) {$2$};
     \node (5) at ($(-0.75cm,0)+(90:1.3cm)$) {$3$};
   \node (3) at (60:1.3cm) {$4$};
     \node (4) at (-60:1.3cm) {$5$};

     \draw[thick,externalline] (-1.5cm,0) -- (1);
      \draw[thick,externalline] (-1.5cm,0) -- (2);
    \draw[thick,externalline] (0,0) -- (3);
    \draw[thick,externalline] (0,0) -- (4);
     \draw[thick,externalline] (-0.75cm,0) -- (5);
     \draw[thick] (0,0)-- +(-1.5cm,0);
     \node (cond2) at (-0.5,-2) {$D+\Delta_3=\Delta_1+\Delta_2+\Delta_4+\Delta_5$} ; 
\end{scope}
\begin{scope}[xshift=9.5cm]
\foreach \i in {1,...,5}
    \draw[thick] (360*\i/5:0.75cm) --(360*\i/5+360/5:0.75cm);
\foreach \i in {1,...,5}
    \draw[thick,externalline] (-360*\i/5:0.75cm) -- +(-360*\i/5:0.6cm) node[pos=1.4] {\i} ;
     \node (cond3) at (0.3,-2) {$2D=\Delta_1+\Delta_2+\Delta_3+\Delta_4+\Delta_5$} ; 
\end{scope}
\end{tikzpicture}
\end{center}
\caption{Five-point graphs for which the remainder in \eq{lr} vanishes.}\label{fig:5pt}
\end{figure}

\paragraph{Case 2: five external points.} For five external points, we find that vanishing of the remainder is only possible if one of several conditions on $\Delta$'s is satisfied. We list them below together with the corresponding graphs that realize each of the conditions:
    \begin{subequations}\label{eq:5PointConditions}
\begin{align}
        &D=\Delta _1+\Delta _2+\Delta _3+\Delta _4+\Delta _5\\
        &2 D=\Delta _1+\Delta _2+\Delta _3+\Delta _4+\Delta _5  
        \\
        &\Delta _1+\Delta _2=\Delta _3+\Delta _4+\Delta _5 \ (+\  \text{symmetries}) \label{eq:5PointConditions3}
        \\
        &D+\Delta _1=\Delta _2+\Delta _3+\Delta _4+\Delta _5 \ (+\  \text{symmetries}) \label{eq:5PointConditions4}
\end{align}
\end{subequations}
where we write ``+ symmetries'' to indicate conditions related to the given one by permutations. A representative of the first case is the cross graph. Some examples representing each of the other cases are given by the graphs on figure \ref{fig:5pt}.
By ''symmetries'' in \eqref{eq:5PointConditions} we mean that the indices can be permuted in a certain way to obtain other possible constraints, like for the four point case. Just as before, not every permutation is allowed because we assumed ordering. In particular, in \eqref{eq:5PointConditions3} the two dimensions on the l.h.s can only be neighboring like $1-2$ or $4-5$. Condition \eqref{eq:5PointConditions4} exists for all choices of the dimension on the l.h.s., in the graphs above we chose it to be $\Delta_3$. Notice that in contrast to the square with four external legs, the pentagon is non-trivial, meaning it does not reduce to a simpler graph with a star-triangle transformation.

    It seems in fact likely that any 5pt Yangian-invariant graph reduces to one of these graphs as they seem to cover the full range of topologies compatible with Yangian constraints.

\paragraph{Case 3: cross-type graphs with any number of legs.} Another case we found when the remainder vanishes is the following: for any number of points we can set $\kappa_{ij}$ to zero and impose
\begin{equation}\label{cco1}
    \sum_{i=1}^N \Delta_{i}= D \, 
\end{equation}
and
\begin{equation}
\label{sco1}
\begin{split}
    s_k-s_1 &= 
      \dfrac{\Delta_1+\Delta_k}{2}+\sum_{j=k+1}^{N} \Delta_j-D =-\dfrac{\Delta_1+\Delta_k}{2}-\sum_{j=2}^{k-1} \Delta_j
\end{split}
      \, ,\quad k=2,\ldots N
\end{equation}
    This is realized, in particular, by the $N$-cross graphs.

\bigskip

This list of cases is certainly not exhaustive, and for instance for 6 external points there are obviously more possibilities -- though we also observe that for certain simple graphs the remainder will not vanish. We hope to pursue a detailed exploration of all these options to the future.

 Below  we review the GKZ equations and their properties, and then we discuss the implications of these observations for some nontrivial Feynman graphs.

\subsection{Review of GKZ systems and $\mathcal{A}$-hypergeometric functions}

Gel'fand-Kapranov-Zelevinsky (GKZ) systems are special systems of differential equations associated to toric actions on manifolds. Their key feature relevant to us is that their properties are very well studied and we can follow known recipes to obtain solutions. The solutions are known as $\mathcal{A}$-hypergeometric functions and are constructed canonically given the system. We will only briefly review the aspects relevant to this work and refer to the original papers \cite{gel1989hypergeometric,Gelfand:1990bua} and reviews \cite{delaCruz:2019skx,Henn:2023tbo} for details.

GKZ systems are defined by the following data:
\begin{itemize}
    \item A $m \times n$ matrix $\mathcal{A}$, also called the toric matrix, with integer entries, such that the vector $\{1,\ldots, 1\}$ lies in its column span. 
    \item A vector $b \in \mathbb{R}^{m}$.  
\end{itemize}
The GKZ system then is the following family of differential equations in variables $z_i$, $i=1\ldots n$ for a function $\Phi(z_1,\dots,z_n)$:
\begin{itemize}
    \item For all $\ell \in \mathbb{Z}^n$ such that $\ell \in \ker(\mathcal{A})$,
        \begin{equation}
            \mathcal{A}  \ell = 0 \ ,
        \end{equation}
one has an equation:
        \begin{equation}\label{eq:GKZtoric}
            \left(\prod_{\ell_i >0 } \partial_{z_i}^{\ell_i} - \prod_{\ell_i <0 } \partial_{z_i}^{-\ell_i}\right)\Phi=0
        \end{equation}
\item The scaling equations 
        \begin{equation}\label{eq:GKZscale}
           \left( \sum_j \mathcal{A}_{ij}z_j \dfrac{\partial}{\partial z_j} - b_i\right)\Phi=0
        \end{equation}
\end{itemize}
Though the definition involves an infinite number of differential operators for all $\ell$, there are only a finite number of them generating the relevant $D$-module, corresponding to a basis in $\ker(\mathcal{A})$ \cite{gel1989hypergeometric}.

At this stage it is useful to note that in our case the $z_i$ variables will be the Poincare invariants $x_{ij}^2$, the matrix ${\cal A}$ will have entries equal to 0 or 1, and the vector $b$ will contain the external scaling dimensions $\Delta_i$.

A simple example (discussed e.g. in \cite{gel1989hypergeometric}) would be the GKZ system for $m=2,n=3$ associated to the matrix:
\begin{equation}
    \mathcal{A} = \begin{pmatrix}
        1& 2 & 0 \\
        1 & 0& 2 \\
    \end{pmatrix}
\end{equation}
The kernel is then spanned by a single vector:
\begin{equation}
    \ker(A)=\mathbb{Z} \cdot \begin{pmatrix}
        2\\
        -1\\
        -1
    \end{pmatrix}
\end{equation}
And hence the independent differential equations are given by:
\begin{equation}\label{eq:GKZ:2-1-1}
\begin{split}
        &\left(\dfrac{\partial^2}{\partial z_1^2} - \dfrac{\partial^2}{\partial z_2 \partial z_3}\right)\Phi=0
        \\
        &\left(z_1 \partial_{z_1} + 2z_2\partial z_2 -b_1\right)\Phi=0\\
        &\left(z_1 \partial_{z_1}+2z_3 \partial z_3 - b_2\right)\Phi=0
\end{split}
\end{equation}

Let us explain how these equations can be solved. First, solving the scaling equations, we get: 
\begin{equation}
    \Phi(z_1,z_2,z_3)=z_2^{\frac{b_1}{2}}z_3^{\frac{b_2}{2}} \Phi_0\left(  \dfrac{z_1^2}{z_2 z_3}\right)
\end{equation}
Then the first equation in \eqref{eq:GKZ:2-1-1} becomes a second order hypergeometric equation in $x=\frac{z_1^2}{z_2 z_3}$, which is solved by:
\begin{equation}\label{eq:GKZ2-1-1solutions}
    \Phi_0\left(x=  \dfrac{z_1^2}{z_2 z_3}\right) = c_1
   \, _2F_1\left(-\frac{b_1}{2},-\frac{b_2}{2};\frac{1}{2};\frac{x}{4}\right)+  c_2 \sqrt{x} \,
   _2F_1\left(\frac{1}{2}-\frac{b_1}{2},\frac{1}{2}-\frac{b_2}{2};\frac{3}{2};\frac{x}{4}\right)
\end{equation}

Many properties of general GKZ systems are known.  In particular GKZ systems are holonomic, meaning that they have a finite dimensional space of solutions. The solutions themselves are given by $\Gamma$-series or, more specifically, $\mathcal{A}$-hypergeometric functions.
Furthermore, for any vector $\gamma$ such that
\begin{equation}
    \sum_{j} \mathcal{A}_{ij} \gamma_j = b_i
\end{equation}
one can show that the series
\begin{equation}\label{eq:GammaSeries}
    \Phi_{\gamma}(z) = \sum_{\ell \in \ker{\mathcal{A}} }  \dfrac{z^{\gamma+\ell}}{\prod\limits_{i=1}^n\Gamma\left( \gamma_i+l_i+1 \right) }
\end{equation}
is a formal solution to the equations  (\ref{eq:GKZtoric}), (\ref{eq:GKZscale}).  
E.g. in our example \eqref{eq:GKZ:2-1-1}, the first basis solution in \eqref{eq:GKZ2-1-1solutions} corresponds to the choice: $\gamma= \begin{pmatrix}
    0\\
    b_1/2\\
    b_2/2\\
\end{pmatrix}$.

Notice that  \eq{eq:GammaSeries} gives a series infinite in both direction for generic values of $\gamma$. To get the actual solutions one more step is needed, which essentially consists of choosing enough of the $\gamma$'s in such a way that the series terminates in one direction, due to poles of the $\Gamma$-functions. The number of ways to do that gives the number of solutions.  This can be formalized in several ways. One, initially presented in \cite{gel1989hypergeometric}, solves the problem in terms of the Newton polytope associated to the column vectors of $\mathcal{A}$. For generic values of the $b$-parameters, the number of solutions is given by the volume of this polytope. Next, one chooses a triangulation of the Newton polytope, and associates a $\Gamma$-series of the form \eqref{eq:GammaSeries} to each simplex in the triangulation, hence creating a basis in the space of solutions. Certain conditions called regularity and unimodularity should be satisfied by the triangulation, however, we do not want to go into detail here and refer to the literature \cite{saito2013grobner, Ananthanarayan:2022ntm,Stienstra:2005nr,delaCruz:2019skx}. Some more care is also needed in the case of special parameters $b$.

For general GKZ systems, these methods were developed within algebraic analysis \cite{saito2013grobner,sattelberger2019d}, and multiple software solutions to deal with GKZ systems and produce bases along with triangulations are known \cite{Ananthanarayan:2022ntm}.
\\

\subsection{The Yangian GKZ systems}

Now that we have described GKZ equations in general, let us proceed to identifying the Yangian system with a concrete type of these equations.

Recall that in section \ref{sec:ycons} we argued that all Yangian symmetric integrals (up to trivial factors of the kind $x_{ij}^{2\kappa_{ij}}$) are annihilated by the $\LL_{ijkl}$ operators given in \eqref{eq:LLconjugate}. Notice that these operators
in fact look like a part of some GKZ system. Indeed, in \cite{Pal:2023kgu} they were shown to correspond to a very special type of GKZ systems.  While in \cite{Pal:2023kgu} the authors considered the relation to GKZ specifically for the $N$-cross integrals, here we identified GKZ operators inside Yangian equations for generic Feynman graphs. Moreover, as discussed in section \ref{sec:ycons} for some graphs the Yangian will reduce precisely to the GKZ equations.

Let us describe the relevant GKZ system explicitly (we mostly follow the notation of \cite{Pal:2023kgu}). We can read it off from the explicit form of the operators \eqref{eq:LLconjugate}. It is written in terms of variables
\begin{equation}\label{eq:vijvars}
    v_{ij}=x^2_{ij}
\end{equation}
which correspond to the set of $z$-variables in our review above. Next, the toric matrix is the following $N \times \dfrac{N(N-1)}{2}$ matrix:
\begin{equation}\label{eq:YangianToricMatrices}
\mathcal{A}_{i,jk}=\delta_{ij}+\delta_{ik}
\end{equation}
and $b_i =-\Delta_i$. Denote by $l_{ij}$ the vectors which span the kernel of the ${\cal A}$-matrix:
\begin{equation}\label{eq:YangianKerA}
    \sum_{j<k}\mathcal{A}_{i,jk} l_{ij} = 0.
\end{equation}
Then the corresponding GKZ operators are as follows:
\begin{equation}\label{eq:YangianGKZ}
    \begin{split}
        &\prod_{l_{ij}>0} \dfrac{\partial^{l_{ij}}}{\partial v_{ij}^{ l_{ij}}} -  \prod_{l_{ij}<0} \dfrac{\partial^{-l_{ij}}}{\partial v_{ij}^{-l_{ij}}} 
        \\
        &\sum_{j<k}\mathcal{A}_{i,jk} v_{jk}\dfrac{\partial}{\partial v_{jk}} + \Delta_i
    \end{split}
\end{equation}
\\
According to the discussion above the solutions to the GKZ systems will take the form of $\Gamma$-series \eqref{eq:GammaSeries}. The summation is over the  kernel of $\mathcal{A}$ with $l_{ij}$ being the exponent of $v_{ij}$ in this case. Notice that the condition \eq{eq:YangianKerA} for $l_{ij}$ is the same as for $\alpha_{ij}$ in the definition of cross-ratios. Hence, for any given choice of cross ratios, $l_{ij}$ are linearly expressed in terms of $\alpha_{ij}^A$. Similarly to the counting of cross ratios, this also means that the dimension of the kernel of $\mathcal{A}$ is $N_{cr}$.
Furthermore, the exponents $\gamma$ in \eqref{eq:GammaSeries} are precisely the exponents $\beta_{ij}$ in the conformal prefactor in \eqref{eq:SystemForBeta}. Therefore we clearly see that the solutions to the GKZ system are indeed, as expected, a series in cross-ratio variables times a conformal weight prefactor.
\\

Recall that in terms of variables \eqref{eq:vijvars} the operators $ \LL_{ijkl}$ have the form:
\begin{equation}
     \LL_{ijkl}=\dfrac{\partial^2}{\partial v_{ij}\partial v_{kl}} - \dfrac{\partial^2}{\partial v_{ik}\partial v_{jl}}\,.
\end{equation}
These can be presented as in \eqref{eq:YangianGKZ} by choosing
\begin{equation}
    l_{ij} = l_{kl} =1 \, ,\quad l_{ik}=l_{jl}=-1
\end{equation}
These satisfy \eqref{eq:YangianKerA} and hence operators $\LL_{ijkl}$ lie in the GKZ ideal.  The scaling equations in \eqref{eq:YangianGKZ} are nothing but conformal invariance conditions, specifically, the dilatation operator, rewritten in variables $v_{ij}$. Hence, the full GKZ system should be equivalent to the set of constraints imposed by conformal symmetry and the level one momentum generator.  
For concreteness, in appendix \ref{app:gkzex} we also provide a few explicit examples of these GKZ systems arising from Yangian equations.

Let us point out that these particular GKZ systems were partially investigated already in one of the original GKZ papers 
\cite{gel1989hypergeometric}, see example 3.3.2. Those results also establish the useful property that the GKZ ideal is generated by the $\LL_{ijkl}$ operators. Note that in that work also another similar system is referred to as corresponding to Grassmanian manifolds (example 3.3.1), and in that case the full set of solutions is even listed explicitly. While the authors mention that  an analogous solutions list is not known for our GKZ system, one may hope that it could be within reach (see some later developments in \cite{gel1996hypergeometric,gel1991hypergeometric}). A result of this kind would thus provide an explicit basis of functions whose combination gives various nontrivial Yangian invariant Feynman integrals.

\subsection{Comments and implications for Feynman graphs}
\label{sec:impl}
Let us emphasize one key point -- namely, the GKZ equations are very constraining and are expected to only have a finite-dimensional space of solutions. Thus up to a conjugation by propagators the graphs satisfying the GKZ system will be expressed in terms of the same basis functions as the $N$-cross graph. Moreover, this space of functions is given by the well-controlled space of $\mathcal{A}$-hypergeometric systems, for the toric matrices that we write out. Of course the coefficients in the linear combination will be different for different graphs, determined essentially by their symmetries and other boundary conditions as usual in the Yangian bootstrap program \cite{Loebbert:2019vcj}.

In particular, in section \ref{sec:ycons} we established that for the pentagon graph in figure \ref{fig:5pt} the Yangian equations reduce precisely to the GKZ system. Accordingly, we conclude that this Feynman integral will be expressed as a linear combination of the \textit{same} hypergeometric functions that form the basis for the 5-cross integral. Notice that although the pentagon graph does have many trivalent vertices, repeated application of star-triangle identities does not seem to reduce it to any standard simple integral. Yet our GKZ  analysis suggests which basis of special functions it should be expressed in. It would be very interesting to verify this directly by computing the integral, as well as to look for more examples where such statements can be made.

Finally let us also highlight that our analysis which allows one to reduce the Yangian equations to the GKZ system for a particular graph relies on several steps. Namely, it is important that the ${\cal L}_{ik}$ operators reduce to individual $\hat L_{ijkl}$, and moreover we impose that the remainder terms in \eq{lr} vanish. We have found this is true for a range of examples, and while we do not expect this to be a general property, it is likely to hold for a large class of graphs. It would be important in the future to look for more examples of this kind and study them in detail.

\section{Conclusions}

In this paper we derived the general form of Yangian equations in invariant conformal variables, as well as elucidating their link with GKZ systems in high generality. Having these general and rather widely applicable proposals opens the way to exploring a significant range of directions in the future. Let us list some of them.

\begin{itemize}
    \item While highly general, our derivation of the Yangian equations in cross-ratios in section 4 assumed that the dimension is not too small compared to the number of external legs. It would be important to explore how to extend the derivation to the completely general case. One promising avenue is to invoke an analyticity argument in the dimension as the latter mostly enters as just a parameter in our equations, and then look for an appropriate reduction of the system to take into account possible dependencies of the variables.
    
    \item Another important further step is to clarify when the remainder functions that distinguish Yangian operators from GKZ ones vanish, and when do the sums of GKZ operators reduce to individual ones -- both points discussed preliminarily in section 5. Addressing the latter point is also closely related to one more key question of rigorously establishing the functional dependencies between Yangian and/or GKZ operators, as well as the number of independent solutions to the systems. Advances in this direction can likely come from employing the powerful algebraic machinery developed for describing GKZ differential ideals and more generally D-modules.

\item With a clearer picture of when Yangian equations exactly reduce to GKZ one should be able to establish a likely large class of Feynman graphs whose calculation will be within reach by reducing them to a linear basis of hypergeometric solutions to GKZ equations. The coefficients in the combination will then be fixed by boundary conditions and symmetries as is typical for the Yangian bootstrap. As a first step it would already be very interesting to try and compute this way the pentagon with 5 external legs from figure \ref{fig:5pt} for which as we established the basis functions should be, remarkably, the same as for the 5-cross. 

\item One more natural question is to try and exploit the rewriting of Yangian equations as 'GKZ terms plus a remainder with 1st order derivatives' even when the remainder does not vanish.

    \item As we mentioned, typically the GKZ systems are associated to Feynman integrals via the  Lee-Pomeransky  \cite{Lee:2013hzt} representation of momentum space integrals. In this scenario, to obtain the GKZ system the variables of the integral should be extended -- one should make all the coefficients of the Lee-Pomerasky polynomial generic. The true integral is then recovered by a rather complicated reduction. In contrast, in our case, the GKZ system is written directly in the initial variables at hand -- the Poincare invariants of external points. Therefore no reduction is needed and the Yangian invariant Feynman integrals are exactly $\mathcal{A}$-hypergeometric functions of conformal cross-ratios. It would be interesting to understand the interplay between our approach and the one based on Lee-Pomeransky representation more generally.
    \item The GKZ systems are simpler at generic values of $\Delta_i$ and $D$. If one sets these to certain integer parameters, one should take limits carefully. This is similar to the standard issues of GKZ systems arising for momentum space integrals and would be interesting to explore in our context.
    
    \item One more interesting reduction would correspond to bringing together some external legs of the graph, in order e.g. to try to reach Basso-Dixon graphs starting from the Yangian invariant case. It may be possible to do it at least in some cases by using advanced properties of solutions to GKZ systems. 
    \item As we already mentioned, the Yangian differential equations in $D=2$ were recently identified with Picard-Fuchs equations for Calabi-Yau manifolds, while the integrals themselves were computed in terms of periods \cite{Duhr:2022pch,Duhr:2024hjf}. Now that we established a more precise relation between the Yangian equations and GKZ systems, one could ask whether this has anything to do with extending the Calabi-Yau geometry to generic $D$. This is especially interesting since GKZ systems do appear in the context of mirror symmetry as equations for Calabi-Yau periods \cite{Hosono:1995bm}.

\item Yangian symmetry essentially implies that the Feynman graphs we study are eigenstates of integrable spin chains based on the conformal group. It would be important to study the application of recently developed separation of variables (SoV) methods for correlators in higher-rank spin chains \cite{Cavaglia:2019pow,Gromov:2019wmz,Maillet:2020ykb,Gromov:2022waj} to this case as already initiated in \cite{Cavaglia:2021mft,Cavaglia:2018lxi}, with a view of obtaining new insights for these Feynman integrals, as well as to investigate the interplay between SoV and GKZ systems.

\item There are curious links between Yangian symmetry or GKZ equations and other questions in CFT: in \cite{Pal:2023kgu} relations were proposed with conformal block/Casimir equations, in \cite{Rigatos:2022eos} with holographic correlators, in \cite{Gelfond:2008ur,Vasiliev:2001dc} 
 with higher spin theories where GKZ-like equations come up, and let us also mention the embedding space formalism where some of the equations may simplify. These deserve to be explored further. 

\item In addition to the conformal case, Yangian invariance has been established for some massive Feynman integrals as well \cite{Loebbert:2020glj,Loebbert:2020tje,Loebbert:2020hxk}. It would be interesting to try to extend our results to this setting.

\end{itemize}

\section*{Acknowledgements}
\noindent We thank P.~Anempodistov, B.~Basso, A.~Gorsky, V.~Kazakov, F.~Loebbert,
A.~Morozov, E.~Olivucci, K.~Ray, V.~Rubtsov, M.~Vasiliev and  Y.~Zenkevich  for insightful discussions. We are also grateful to the Bethe Centre for Theoretical Physics in Bonn for hospitality during the workshop `Fishnets: Conformal Field Theories and Feynman Graphs' in September 2024 where preliminary results of this project were presented. Nordita is supported in part by NordForsk.

\appendix

\section{Details on geometric constraints for the Loom}
\label{app:geom}

Here we argue that if the conditions \eq{con1}, \eq{con2} are satisfied, then at least locally the graph can be drawn on the Loom.

\begin{figure}[H]
    \centering
    \includegraphics[scale=0.6]{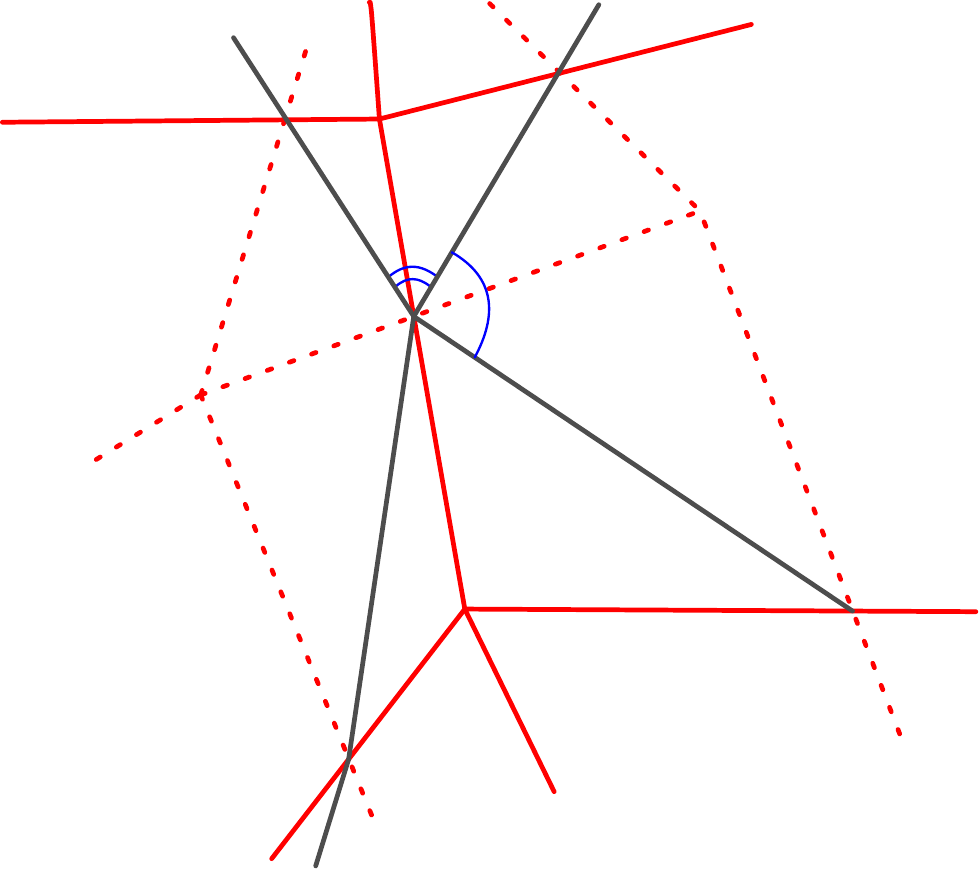}
    \caption{Local Loom reconstruction. A portion of the graph is drawn by red lines, and it's dual by dashed lines. The gray segments are intentionally drawn as if they don't form an intersection of two straight lines. However, the sum of the two angles highlighted in blue turns out to give $\pi$.}
    \label{fig:LocalLoom}
\end{figure}

We will argue that the Baxter lattice of straight lines of the Loom construction can be built ''around'' the graph. 
Let a graph satisfy conformal and dual conformal conditions. This means that the sum of dimensions at each vertex of both the graph and its dual are equal to $D$. The first condition -- conformal invariance --  means that we consistently identify the propagator powers with angles by \eq{da}. That is, at each intersection of the Feynman graph and its dual we construct four outgoing lines, going from one intersection to the other. Locally this looks as shown on figure \ref{fig:LocalLoom}. 
The conformality condition ensures that the geometric constraint of the sum of all angles of a black $n$-gon cell being $\pi(n-2)$ is satisfied. At this stage we just have a collection of lines going from one intersection of the graph and its dual to the other.\footnote{Notice that so far the argument goes through even for the hexa-star graph from figure \ref{fig:6ptKagome}, which however cannot be drawn on the Loom as the next step will fail.} 

Next, we impose the dual symmetry condition \eq{con2}. This means that the dimensions $\Delta$ of the dual graph can also be mapped to angles through which the lines of the dual graph go. Finally, we use that dimensions of the original and dual line are related as:
\begin{equation}
    \Delta^{(\text{dual})} = \dfrac{D}{2}-\Delta
\end{equation}
Therefore the two highlighted angles in fig. \ref{fig:LocalLoom} add up to $\pi$ and hence the four outgoing segments are really two intersecting lines. This is true at every vertex. Hence we get a Baxter lattice made up from straight lines.

The argument we discussed works locally for internal vertices. Some additional care should be taken for external vertices (e.g. to avoid the problem of having 'too many legs' as for the graph on figure \ref{fig:trlegs} which cannot be drawn on the Loom), and to ensure that everything works globally. We leave a more detailed exploration of this to the future.

\section{The double-star integral and the star-triangle transformation}\label{sec:AppendixDoubleStarIntegral}

Recall that the star-triangle transformation is the identity between a 3-leg integral and a triangle graph:

\begin{equation}
    \int \dfrac{d^D x_0}{x_{10}^{2\Delta_1}x_{20}^{2\Delta_2}x_{30}^{2\Delta_3}}=\vcenter{\hbox{\StarGraph}}= \vcenter{\hbox{\TriangleGraph}} = \dfrac{a(\Delta_1,\Delta_2,\Delta_3)}{x_{12}^{D-2\Delta_3}x_{23}^{D-2\Delta_1} x_{13}^{D-2\Delta_2}}.
\end{equation}
Here $a(\Delta_1,\Delta_2,\Delta_3)= \pi^{D/2}\dfrac{\Gamma_{D/2-\Delta_1} \Gamma_{D/2-\Delta_2} \Gamma_{D/2-\Delta_3}}{\Gamma_{\Delta_1}\Gamma_{\Delta_2}\Gamma_{\Delta_3}}$. We remind that we use the notation: $\Gamma_x=\Gamma(x)$.

Let us now consider the following two graphs:

\begin{center}
   \begin{tikzpicture}
    \node (1) at ($(-1cm,0)+(-90-30:1.3cm)$) {$1$};
     \node (2) at ($(-1cm,0)+(90+30:1.3cm)$) {$2$};
   \node (3) at (60:1.3cm) {$3$};
    \node (4) at (-60:1.3cm) {$4$};
     \draw[thick,externalline] (-1cm,0) -- (1);
      \draw[thick,externalline] (-1cm,0) -- (2);
    \draw[thick,externalline] (0,0) -- (3);
    \draw[thick,externalline] (0,0) -- (4);
        \draw[thick] (0,0)-- +(-1cm,0);

    \node (11) at ($(4,0)+(-90-45:1.3cm)$) {$1$};
    \node (12) at ($(4,0)+(90+45:1.3cm)$) {$2$};
    \node (13) at ($(4,0)+(45:1.3cm)$) {$3$};
    \node (14) at ($(4,0)+(-45:1.3cm)$) {$4$};
   
    \draw[thick,externalline] (4,0) -- ($(4,0)+(45:1cm)$);
    \draw[thick,externalline] (4,0) -- ($(4,0)+(-45:1cm)$);
    \draw[thick,externalline] (4,0) -- ($(4,0)+(90+45:1cm)$);
    \draw[thick,externalline] (4,0) -- ($(4,0)+(-90-45:1cm)$);
     \draw[thick,shorten <=2pt,shorten >=2pt] ($(4,0)+(-1pt,0)+(-45:1cm)$) -- ($(4,0)+(-1pt,0)+(45:1cm)$);
\end{tikzpicture} 
\end{center}
The graph on the left can be transformed into the one on the right via the star-triangle transformation. The conformal conditions for the integral read:
\begin{equation}\label{eq:ConditionsDoubleStar}
    \Delta_1+\Delta_2 = D-\Delta_0 = \Delta_3+\Delta_4\,.
\end{equation}
where $\Delta_0$ is the dimension of the internal propagator.

Applying star-triangle 
to the vertex connected to $x_3$ and $x_4$, we get a cross integral with a propagator attached to point 3 and 4:
\begin{equation}\label{eq:DoubleStarTocross}
    I_{\scalebox{0.15}{\doublestar}}(\mathbf{x}|\Delta_1,\Delta_2,\Delta_3,\Delta_4) =  a(\Delta_3,\Delta_4,\Delta_0)  \dfrac{1}{x_{34}^{D-2\Delta_0}}  I_{\boldsymbol{+}}\left(\mathbf{x} \Big|\Delta_1,\Delta_2,\dfrac{D}{2}-\Delta_4,\dfrac{D}{2}-\Delta_3 \right)
\end{equation}
The dimension $\Delta_0$ can be expressed in terms of the external ones due to conformal conditions, hence the power of the propagator in the prefactor is  $D-2\Delta_0=-D+2\Delta_1+2\Delta_2$

Since we have used the star-triangle transformation, conformal symmetry is not broken. Therefore the parameters of resulting cross integral will satisfy the condition that the sum of propagator powers is $D$. Indeed it is a simple check that the sum of dimensions in the r.h.s of \eqref{eq:DoubleStarTocross} is $D$ given the condition \eqref{eq:ConditionsDoubleStar}.

Referring to the discussion of operators \eqref{eq:LLconjugate}, notice that the cross integral clearly satisfies GKZ equations with the usual $\LL_{ijkl}$ operators. From this we clearly see that the double-star integral would be annihilated by the conjugated operators, with  $\kappa_{34} = -D/2+\Delta_1+\Delta_2$
and the rest of $\kappa_{ij}$ vanishing.
\\

\section{Results of bootstrapping the cross and Appell series.}\label{sec:AppendixCrossAndAppel}
Here we collect the results of \cite{Loebbert:2019vcj} on the cross integral. The integral is defined as
\beq
    I_{\boldsymbol{+}}=\int \frac{d^Dx_0}{x_{10}^{2\Delta_1}x_{20}^{2\Delta_2}x_{30}^{2\Delta_3}x_{40}^{2\Delta_4}}
\eeq
with $\Delta_1+\Delta_2+\Delta_3+\Delta_4=D$. Then one has

\begin{equation}
 I_{\boldsymbol{+}}\left(\mathbf{x} \Big|\Delta_1,\Delta_2,\Delta_3,\Delta_4 \right) = V_4 g(u,v)
 \end{equation}
 where $u,v$ are conformal cross-ratios. 
The prefactor is given as:
\begin{equation}
   V_4=x_{24}^{-2 \Delta _2} x_{13}^{2 \Delta _4-D}
   x_{14}^{-D+2 \Delta _2+2 \Delta _3} x_{34}^{D-2
   \Delta _3-2 \Delta _4}
\end{equation}
The four basis solutions in are chosen as:
\begin{equation}
 \begin{split}
 g_1=&F_4\left(\Delta _2,\frac{D}{2}-\Delta _4,\frac{D}{2}-\Delta _3-\Delta _4+1,-\frac{D}{2}+\Delta _2+\Delta _3+1,u,v\right)
   \\
 g_2=&u^{-\frac{D}{2}+\Delta _3+\Delta _4} F_4\left(-\frac{D}{2}+\Delta _2+\Delta _3+\Delta _4,\Delta _3,-\frac{D}{2}+\Delta
   _3+\Delta _4+1,-\frac{D}{2}+\Delta _2+\Delta _3+1,u,v\right) \\
 g_3=&v^{\frac{D}{2}-\Delta _2-\Delta _3} F_4\left(\frac{D}{2}-\Delta _3,D-\Delta _2-\Delta _3-\Delta _4,\frac{D}{2}-\Delta
   _3-\Delta _4+1,\frac{D}{2}-\Delta _2-\Delta _3+1,u,v\right) \\
 g_4=&u^{-\frac{D}{2}+\Delta _3+\Delta _4} v^{\frac{D}{2}-\Delta _2-\Delta _3} F_4\left(\Delta _4,\frac{D}{2}-\Delta
   _2,-\frac{D}{2}+\Delta _3+\Delta _4+1,\frac{D}{2}-\Delta _2-\Delta _3+1,u,v\right) 
\end{split}
\end{equation}
Up to constants the hypergeometric series in \eqref{eq:YangianGKZsolution} is given by the Appell series:
\begin{equation}
\Phi_1 \sim \Tilde{\Phi}_1 = F_4\left(\frac{D}{2}-\Delta _3,\Delta _2,\frac{D}{2}-\Delta _3-\Delta _4+1,\frac{D}{2}-\Delta _1-\Delta _3+1,\xi^1,\xi^2 \right)
\end{equation}
To relate these solutions, notice that:
\begin{equation}
    \xi^1=\frac{u}{v}\, , \, \xi^2=\dfrac{1}{v} \, ,\quad u = \dfrac{\xi^1}{\xi^2} \, , v=\dfrac{1}{\xi^2}
\end{equation}
Using transformation \cite{appell1926fonctions}:
\begin{equation}
    \begin{aligned}
& F_4(a, b ; c, d ; x, y)= \\
& =\frac{\Gamma(d) \Gamma(b-a)}{\Gamma(d-a) \Gamma(b)}(-y)^{-a} F_4\left(a, a-d+1 ; c, a-b+1 ; \frac{x}{y}, \frac{1}{y}\right)+ \\
&+ \frac{\Gamma(d) \Gamma(a-b)}{\Gamma(d-b) \Gamma(a)}(-y)^{-b} F_4\left(b, b-d+1 ; c, b-a+1 ; \frac{x}{y}, \frac{1}{y}\right)
\end{aligned}
\end{equation}
we get:
\begin{equation}
    \Tilde{\Phi}_1=v^{\Delta_2}\left( (-1)^{\Delta _2}\frac{ \Gamma
   _{\frac{D}{2}-\Delta _1-\Delta _3+1} \Gamma
   _{\frac{D}{2}-\Delta _2-\Delta _3}}{\Gamma
   _{\frac{D}{2}-\Delta _3} \Gamma
   _{\frac{D}{2}-\Delta _1-\Delta _2-\Delta
   _3+1}} \cdot g_1 +  (-1)^{\Delta _3-\frac{D}{2}}
 \frac{ \Gamma
   _{\frac{D}{2}-\Delta _1-\Delta _3+1} \Gamma
   _{-\frac{D}{2}+\Delta _2+\Delta _3}}{\Gamma
   _{1-\Delta _1} \Gamma _{\Delta
   _2}} \cdot g_3 \right)
\end{equation}
The prefactor $v^{\Delta_2}$ is responsible exactly for the different choice of prefactors between us and \cite{Loebbert:2019vcj}.

\section{Examples of GKZ equations from the Yangian}

\label{app:gkzex}

Let us present a couple of explicit examples for illustration. For instance, as we have argued an arbitrary 4-point graph will be given by a linear combination of the basis functions corresponding to the 4-cross. Their explicit form and the corresponding GKZ conventions were given in \cite{Pal:2023kgu} and in our notation they are as follows. The toric matrix reads
\begin{equation}
    \mathcal{A}=\left(\begin{array}{llllll}
1 & 1 & 1 & 0 & 0 & 0 \\
1 & 0 & 0 & 1 & 1 & 0 \\
0 & 1 & 0 & 1 & 0 & 1 \\
0 & 0 & 1 & 0 & 1 & 1
\end{array}\right)
\end{equation}
The resulting basis $\mathcal{A}$-hypergeometric functions are four series in two variables, and e.g. one of them looks like:
\begin{equation}\label{eq:YangianGKZsolution}
    \begin{aligned}
& \Phi_1^{}=x_{12}^{D-2\left(\Delta_1+\Delta_2\right)}x_{13}^{D-2\left(\Delta_1+\Delta_3\right)}x_{14}^{-2 \Delta_4}x_{23}^{2 \Delta_1-D} \times  \\
& \times \sum_{n_1, n_2=0}^{\infty} \frac{\left(\xi^2\right)^{n_1}\left(\xi^1\right)^{n_2}}{n_1! n_2!\Gamma_{1+n_1-\Delta_1-\Delta_3+\frac{D}{2}} \Gamma_{1+n_2-\Delta_1-\Delta_2+\frac{D}{2}} \Gamma_{1-n_1-n_2-\Delta_4} \Gamma_{1-n_1-n_2+\Delta_1-\frac{D}{2}}}
\end{aligned}
\end{equation}
where $\Gamma_x=\Gamma(x)$. The resulting basis of series is related to  the Appell series of \cite{Loebbert:2019vcj} via analytical continuation, see Appendix \ref{sec:AppendixCrossAndAppel}.

As another example, the toric matrix for $N=5$ is:
\begin{equation}
    \mathcal{A}= \left(\begin{array}{llllllllll}
1 & 1 & 1 & 1 & 0 & 0 & 0 & 0 & 0 & 0 \\
1 & 0 & 0 & 0 & 1 & 1 & 1 & 0 & 0 & 0 \\
0 & 1 & 0 & 0 & 1 & 0 & 0 & 1 & 1 & 0 \\
0 & 0 & 1 & 0 & 0 & 1 & 0 & 1 & 0 & 1 \\
0 & 0 & 0 & 1 & 0 & 0 & 1 & 0 & 1 & 1
\end{array}\right)
\end{equation}
The solution space is given by 11 $\Gamma$-series in five cross ratio variables, presented more explicitly in \cite{Pal:2023kgu}.

 \bibliographystyle{JHEP.bst}
 \bibliography{GKZvsY,Loom_FCFT,FeynmanRef}

\providecommand{\href}[2]{#2}\begingroup\raggedright\begin{thebibliography}{10}

\bibitem{Weinzierl:2022eaz}
S.~Weinzierl, \emph{{Feynman Integrals}} (1, 2022),
  \href{https://doi.org/10.1007/978-3-030-99558-4}{10.1007/978-3-030-99558-4},
  [\href{https://arxiv.org/abs/2201.03593}{{\ttfamily 2201.03593}}].

\bibitem{Vanhove:2018mto}
P.~Vanhove, \emph{{Feynman integrals, toric geometry and mirror symmetry}},  in
  \emph{{KMPB Conference}: {Elliptic Integrals, Elliptic Functions and Modular
  Forms in Quantum Field Theory}}, pp.~415--458, 2019,
  \href{https://doi.org/10.1007/978-3-030-04480-0_17}{DOI}
  [\href{https://arxiv.org/abs/1807.11466}{{\ttfamily 1807.11466}}].

\bibitem{Abreu:2022mfk}
S.~Abreu, R.~Britto and C.~Duhr, \emph{{The SAGEX review on scattering
  amplitudes Chapter 3: Mathematical structures in Feynman integrals}},
  \href{https://doi.org/10.1088/1751-8121/ac87de}{\emph{J. Phys. A} {\bfseries
  55} (2022) 443004} [\href{https://arxiv.org/abs/2203.13014}{{\ttfamily
  2203.13014}}].

\bibitem{Blumlein:2022qci}
J.~Bl\"umlein and C.~Schneider, \emph{{The SAGEX review on scattering
  amplitudes Chapter 4: Multi-loop Feynman integrals}},
  \href{https://doi.org/10.1088/1751-8121/ac8086}{\emph{J. Phys. A} {\bfseries
  55} (2022) 443005} [\href{https://arxiv.org/abs/2203.13015}{{\ttfamily
  2203.13015}}].

\bibitem{Bourjaily:2022bwx}
J.L.~Bourjaily et~al., \emph{{Functions Beyond Multiple Polylogarithms for
  Precision Collider Physics}},  in \emph{{Snowmass 2021}}, 3, 2022
  [\href{https://arxiv.org/abs/2203.07088}{{\ttfamily 2203.07088}}].

\bibitem{Beisert:2010jr}
N.~Beisert et~al., \emph{{Review of AdS/CFT Integrability: An Overview}},
  \href{https://doi.org/10.1007/s11005-011-0529-2}{\emph{Lett. Math. Phys.}
  {\bfseries 99} (2012) 3} [\href{https://arxiv.org/abs/1012.3982}{{\ttfamily
  1012.3982}}].

\bibitem{Zamolodchikov:1980mb}
A.B.~Zamolodchikov, \emph{{'Fishnet' diagrams as a completely integrable
  system}}, \href{https://doi.org/10.1016/0370-2693(80)90547-X}{\emph{Phys.
  Lett.} {\bfseries 97B} (1980) 63}.

\bibitem{Gurdogan:2015csr}
O.~G\"{u}rdo\u{g}an and V.~Kazakov, \emph{{New Integrable 4D Quantum Field
  Theories from Strongly Deformed Planar $\mathcal N = $ 4 Supersymmetric
  Yang-Mills Theory}}, \href{https://doi.org/10.1103/PhysRevLett.117.201602,
  10.1103/PhysRevLett.117.259903}{\emph{Phys. Rev. Lett.} {\bfseries 117}
  (2016) 201602} [\href{https://arxiv.org/abs/1512.06704}{{\ttfamily
  1512.06704}}].

\bibitem{Chicherin:2017cns}
D.~Chicherin, V.~Kazakov, F.~Loebbert, D.~Mueller and D.-l.~Zhong,
  \emph{{Yangian Symmetry for Bi-Scalar Loop Amplitudes}},
  \href{https://doi.org/10.1007/JHEP05(2018)003}{\emph{JHEP} {\bfseries 05}
  (2017) 003} [\href{https://arxiv.org/abs/1704.01967}{{\ttfamily
  1704.01967}}].

\bibitem{Chicherin:2017frs}
D.~Chicherin, V.~Kazakov, F.~Loebbert, D.~Mueller and D.-l.~Zhong,
  \emph{{Yangian Symmetry for Fishnet Feynman Graphs}},
  \href{https://doi.org/10.1103/PhysRevD.96.121901}{\emph{Phys. Rev.}
  {\bfseries D96} (2017) 121901}
  [\href{https://arxiv.org/abs/1708.00007}{{\ttfamily 1708.00007}}].

\bibitem{Loebbert:2024qbw}
F.~Loebbert and H.~Mathur, \emph{{The Feyn-Structure of Yangian Symmetry}},
  \href{https://arxiv.org/abs/2410.11936}{{\ttfamily 2410.11936}}.

\bibitem{Duhr:2024hjf}
C.~Duhr, A.~Klemm, F.~Loebbert, C.~Nega and F.~Porkert, \emph{{Geometry from
  integrability: multi-leg fishnet integrals in two dimensions}},
  \href{https://doi.org/10.1007/JHEP07(2024)008}{\emph{JHEP} {\bfseries 07}
  (2024) 008} [\href{https://arxiv.org/abs/2402.19034}{{\ttfamily
  2402.19034}}].

\bibitem{Loebbert:2024fsj}
F.~Loebbert and S.F.~Stawinski, \emph{{Conformal four-point integrals:
  recursive structure, Toda equations and double copy}},
  \href{https://doi.org/10.1007/JHEP11(2024)092}{\emph{JHEP} {\bfseries 11}
  (2024) 092} [\href{https://arxiv.org/abs/2408.15331}{{\ttfamily
  2408.15331}}].

\bibitem{Duhr:2023eld}
C.~Duhr, A.~Klemm, F.~Loebbert, C.~Nega and F.~Porkert, \emph{{The Basso-Dixon
  formula and Calabi-Yau geometry}},
  \href{https://doi.org/10.1007/JHEP03(2024)177}{\emph{JHEP} {\bfseries 03}
  (2024) 177} [\href{https://arxiv.org/abs/2310.08625}{{\ttfamily
  2310.08625}}].

\bibitem{Loebbert:2022nfu}
F.~Loebbert, \emph{{Integrability for Feynman Integrals}},  12, 2022
  [\href{https://arxiv.org/abs/2212.09636}{{\ttfamily 2212.09636}}].

\bibitem{Duhr:2022pch}
C.~Duhr, A.~Klemm, F.~Loebbert, C.~Nega and F.~Porkert, \emph{Yangian-invariant
  fishnet integrals in 2 dimensions as volumes of calabi-yau varieties},
  \href{https://arxiv.org/abs/2209.05291}{{\ttfamily 2209.05291}}.

\bibitem{Corcoran:2021gda}
L.~Corcoran, F.~Loebbert and J.~Miczajka, \emph{Yangian ward identities for
  fishnet four-point integrals},
  \href{https://doi.org/10.1007/JHEP04(2022)131}{\emph{JHEP} {\bfseries 04}
  (2022) 131} [\href{https://arxiv.org/abs/2112.06928}{{\ttfamily
  2112.06928}}].

\bibitem{Loebbert:2021qef}
F.~Loebbert and J.~Miczajka, \emph{{Massive Integrability: From Fishnet
  Theories to Feynman Graphs and Back}},
  \href{https://doi.org/10.22323/1.398.0733}{\emph{PoS} {\bfseries EPS-HEP2021}
  (2022) 733} [\href{https://arxiv.org/abs/2109.11937}{{\ttfamily
  2109.11937}}].

\bibitem{Corcoran:2020epz}
L.~Corcoran, F.~Loebbert, J.~Miczajka and M.~Staudacher, \emph{{Minkowski Box
  from Yangian Bootstrap}},
  \href{https://doi.org/10.1007/JHEP04(2021)160}{\emph{JHEP} {\bfseries 04}
  (2021) 160} [\href{https://arxiv.org/abs/2012.07852}{{\ttfamily
  2012.07852}}].

\bibitem{Loebbert:2020glj}
F.~Loebbert, J.~Miczajka, D.~M\"uller and H.~M\"unkler, \emph{{Yangian
  Bootstrap for Massive Feynman Integrals}},
  \href{https://doi.org/10.21468/SciPostPhys.11.1.010}{\emph{SciPost Phys.}
  {\bfseries 11} (2021) 010}
  [\href{https://arxiv.org/abs/2010.08552}{{\ttfamily 2010.08552}}].

\bibitem{Loebbert:2020tje}
F.~Loebbert and J.~Miczajka, \emph{{Massive Fishnets}},
  \href{https://doi.org/10.1007/JHEP12(2020)197}{\emph{JHEP} {\bfseries 12}
  (2020) 197} [\href{https://arxiv.org/abs/2008.11739}{{\ttfamily
  2008.11739}}].

\bibitem{Loebbert:2020hxk}
F.~Loebbert, J.~Miczajka, D.~M\"uller and H.~M\"unkler, \emph{{Massive
  Conformal Symmetry and Integrability for Feynman Integrals}},
  \href{https://doi.org/10.1103/PhysRevLett.125.091602}{\emph{Phys. Rev. Lett.}
  {\bfseries 125} (2020) 091602}
  [\href{https://arxiv.org/abs/2005.01735}{{\ttfamily 2005.01735}}].

\bibitem{Loebbert:2019vcj}
F.~Loebbert, D.~M\"uller and H.~M\"unkler, \emph{{Yangian Bootstrap for
  Conformal Feynman Integrals}},
  \href{https://doi.org/10.1103/PhysRevD.101.066006}{\emph{Phys. Rev. D}
  {\bfseries 101} (2020) 066006}
  [\href{https://arxiv.org/abs/1912.05561}{{\ttfamily 1912.05561}}].

\bibitem{Chicherin:2022nqq}
D.~Chicherin and G.P.~Korchemsky, \emph{{The SAGEX review on scattering
  amplitudes Chapter 9: Integrability of amplitudes in fishnet theories}},
  \href{https://doi.org/10.1088/1751-8121/ac8c72}{\emph{J. Phys. A} {\bfseries
  55} (2022) 443010} [\href{https://arxiv.org/abs/2203.13020}{{\ttfamily
  2203.13020}}].

\bibitem{Drummond:2009fd}
J.M.~Drummond, J.M.~Henn and J.~Plefka, \emph{{Yangian symmetry of scattering
  amplitudes in N=4 super Yang-Mills theory}},
  \href{https://doi.org/10.1088/1126-6708/2009/05/046}{\emph{JHEP} {\bfseries
  05} (2009) 046} [\href{https://arxiv.org/abs/0902.2987}{{\ttfamily
  0902.2987}}].

\bibitem{Drummond:2008vq}
J.M.~Drummond, J.~Henn, G.P.~Korchemsky and E.~Sokatchev, \emph{{Dual
  superconformal symmetry of scattering amplitudes in N=4 super-Yang-Mills
  theory}}, \href{https://doi.org/10.1016/j.nuclphysb.2009.11.022}{\emph{Nucl.
  Phys. B} {\bfseries 828} (2010) 317}
  [\href{https://arxiv.org/abs/0807.1095}{{\ttfamily 0807.1095}}].

\bibitem{Beisert:2010gn}
N.~Beisert, J.~Henn, T.~McLoughlin and J.~Plefka, \emph{{One-Loop
  Superconformal and Yangian Symmetries of Scattering Amplitudes in N=4 Super
  Yang-Mills}}, \href{https://doi.org/10.1007/JHEP04(2010)085}{\emph{JHEP}
  {\bfseries 04} (2010) 085} [\href{https://arxiv.org/abs/1002.1733}{{\ttfamily
  1002.1733}}].

\bibitem{Arkani-Hamed:2012zlh}
N.~Arkani-Hamed, J.L.~Bourjaily, F.~Cachazo, A.B.~Goncharov, A.~Postnikov and
  J.~Trnka, \emph{{Grassmannian Geometry of Scattering Amplitudes}}, Cambridge
  University Press (4, 2016),
  \href{https://doi.org/10.1017/CBO9781316091548}{10.1017/CBO9781316091548},
  [\href{https://arxiv.org/abs/1212.5605}{{\ttfamily 1212.5605}}].

\bibitem{Huang:2010qy}
Y.-t.~Huang and A.E.~Lipstein, \emph{{Dual Superconformal Symmetry of N=6
  Chern-Simons Theory}},
  \href{https://doi.org/10.1007/JHEP11(2010)076}{\emph{JHEP} {\bfseries 11}
  (2010) 076} [\href{https://arxiv.org/abs/1008.0041}{{\ttfamily 1008.0041}}].

\bibitem{Bargheer:2010hn}
T.~Bargheer, F.~Loebbert and C.~Meneghelli, \emph{{Symmetries of Tree-level
  Scattering Amplitudes in N=6 Superconformal Chern-Simons Theory}},
  \href{https://doi.org/10.1103/PhysRevD.82.045016}{\emph{Phys. Rev. D}
  {\bfseries 82} (2010) 045016}
  [\href{https://arxiv.org/abs/1003.6120}{{\ttfamily 1003.6120}}].

\bibitem{Beisert:2017pnr}
N.~Beisert, A.~Garus and M.~Rosso, \emph{{Yangian Symmetry and Integrability of
  Planar N=4 Supersymmetric Yang-Mills Theory}},
  \href{https://doi.org/10.1103/PhysRevLett.118.141603}{\emph{Phys. Rev. Lett.}
  {\bfseries 118} (2017) 141603}
  [\href{https://arxiv.org/abs/1701.09162}{{\ttfamily 1701.09162}}].

\bibitem{Kazakov:2023nyu}
V.~Kazakov, F.~Levkovich-Maslyuk and V.~Mishnyakov, \emph{{Integrable Feynman
  Graphs and Yangian Symmetry on the Loom}},
  \href{https://arxiv.org/abs/2304.04654}{{\ttfamily 2304.04654}}.

\bibitem{Kazakov:2022dbd}
V.~Kazakov and E.~Olivucci, \emph{{The loom for general fishnet CFTs}},
  \href{https://doi.org/10.1007/JHEP06(2023)041}{\emph{JHEP} {\bfseries 06}
  (2023) 041} [\href{https://arxiv.org/abs/2212.09732}{{\ttfamily
  2212.09732}}].

\bibitem{Kazakov:2018gcy}
V.~Kazakov, E.~Olivucci and M.~Preti, \emph{{Generalized fishnets and exact
  four-point correlators in chiral CFT$_{4}$}},
  \href{https://doi.org/10.1007/JHEP06(2019)078}{\emph{JHEP} {\bfseries 06}
  (2019) 078} [\href{https://arxiv.org/abs/1901.00011}{{\ttfamily
  1901.00011}}].

\bibitem{Derkachov:2018rot}
S.~Derkachov, V.~Kazakov and E.~Olivucci, \emph{{Basso-Dixon Correlators in
  Two-Dimensional Fishnet CFT}},
  \href{https://arxiv.org/abs/1811.10623}{{\ttfamily 1811.10623}}.

\bibitem{Alfimov:2023vev}
M.~Alfimov, G.~Ferrando, V.~Kazakov and E.~Olivucci, \emph{{Checkerboard CFT}},
   \href{https://arxiv.org/abs/2311.01437}{{\ttfamily 2311.01437}}.

\bibitem{Bezuglov:2020tff}
M.~Bezuglov, \emph{{Calculation of master integrals in terms of elliptic
  multiple polylogarithms}},
  \href{https://doi.org/10.1142/S0217751X20500633}{\emph{Int. J. Mod. Phys. A}
  {\bfseries 35} (2020) 2050063}
  [\href{https://arxiv.org/abs/2003.05367}{{\ttfamily 2003.05367}}].

\bibitem{Kotikov:2021tai}
A.V.~Kotikov, \emph{{Differential Equations and Feynman Integrals}},  in
  \emph{{Antidifferentiation and the Calculation of Feynman Amplitudes}}, 2,
  2021, \href{https://doi.org/10.1007/978-3-030-80219-6_10}{DOI}
  [\href{https://arxiv.org/abs/2102.07424}{{\ttfamily 2102.07424}}].

\bibitem{Abreu:2020jxa}
S.~Abreu, H.~Ita, F.~Moriello, B.~Page, W.~Tschernow and M.~Zeng,
  \emph{{Two-Loop Integrals for Planar Five-Point One-Mass Processes}},
  \href{https://doi.org/10.1007/JHEP11(2020)117}{\emph{JHEP} {\bfseries 11}
  (2020) 117} [\href{https://arxiv.org/abs/2005.04195}{{\ttfamily
  2005.04195}}].

\bibitem{Bonisch:2021yfw}
K.~B\"onisch, C.~Duhr, F.~Fischbach, A.~Klemm and C.~Nega, \emph{{Feynman
  integrals in dimensional regularization and extensions of Calabi-Yau
  motives}}, \href{https://doi.org/10.1007/JHEP09(2022)156}{\emph{JHEP}
  {\bfseries 09} (2022) 156}
  [\href{https://arxiv.org/abs/2108.05310}{{\ttfamily 2108.05310}}].

\bibitem{Lairez:2022zkj}
P.~Lairez and P.~Vanhove, \emph{{Algorithms for minimal
  Picard\textendash{}Fuchs operators of Feynman integrals}},
  \href{https://doi.org/10.1007/s11005-023-01661-3}{\emph{Lett. Math. Phys.}
  {\bfseries 113} (2023) 37}
  [\href{https://arxiv.org/abs/2209.10962}{{\ttfamily 2209.10962}}].

\bibitem{delaCruz:2024xit}
L.~de~la Cruz and P.~Vanhove, \emph{{Algorithm for differential equations for
  Feynman integrals in general dimensions}},
  \href{https://doi.org/10.1007/s11005-024-01832-w}{\emph{Lett. Math. Phys.}
  {\bfseries 114} (2024) 89}
  [\href{https://arxiv.org/abs/2401.09908}{{\ttfamily 2401.09908}}].

\bibitem{gel1989hypergeometric}
I.M.~Gel'fand, A.V.~Zelevinskii and M.M.~Kapranov, \emph{Hypergeometric
  functions and toral manifolds}, {\emph{Funktsional'nyi Analiz i ego
  Prilozheniya} {\bfseries 23} (1989) 12}.

\bibitem{Gelfand:1990bua}
I.M.~Gelfand, M.M.~Kapranov and A.V.~Zelevinsky, \emph{{Generalized Euler
  integrals and A-hypergeometric functions }},
  \href{https://doi.org/10.1016/0001-8708(90)90048-R}{\emph{Adv. Math.}
  {\bfseries 84} (1990) 255}.

\bibitem{sturmfels2000solving}
B.~Sturmfels, \emph{Solving algebraic equations in terms of a-hypergeometric
  series}, {\emph{Discrete Mathematics} {\bfseries 210} (2000) 171}.

\bibitem{Morozov:2009fc}
A.~Morozov and S.~Shakirov, \emph{{Introduction to Integral Discriminants}},
  \href{https://doi.org/10.1088/1126-6708/2009/12/002}{\emph{JHEP} {\bfseries
  12} (2009) 002} [\href{https://arxiv.org/abs/0903.2595}{{\ttfamily
  0903.2595}}].

\bibitem{Hosono:1995bm}
S.~Hosono, B.H.~Lian and S.-T.~Yau, \emph{{GKZ generalized hypergeometric
  systems in mirror symmetry of Calabi-Yau hypersurfaces}},
  \href{https://doi.org/10.1007/BF02506417}{\emph{Commun. Math. Phys.}
  {\bfseries 182} (1996) 535}
  [\href{https://arxiv.org/abs/alg-geom/9511001}{{\ttfamily
  alg-geom/9511001}}].

\bibitem{batyrev1993variations}
V.V.~Batyrev, \emph{Variations of the mixed hodge structure of affine
  hypersurfaces in algebraic tori}, .

\bibitem{nasrollahpoursamami2016periods}
E.~Nasrollahpoursamami, \emph{Periods of feynman diagrams and gkz d-modules},
  {\emph{arXiv preprint arXiv:1605.04970} (2016) }.

\bibitem{delaCruz:2019skx}
L.~de~la Cruz, \emph{{Feynman integrals as A-hypergeometric functions}},
  \href{https://doi.org/10.1007/JHEP12(2019)123}{\emph{JHEP} {\bfseries 12}
  (2019) 123} [\href{https://arxiv.org/abs/1907.00507}{{\ttfamily
  1907.00507}}].

\bibitem{Grimm:2024tbg}
T.W.~Grimm and A.~Hoefnagels, \emph{{Reductions of GKZ Systems and Applications
  to Cosmological Correlators}},
  \href{https://arxiv.org/abs/2409.13815}{{\ttfamily 2409.13815}}.

\bibitem{Pal:2023kgu}
A.~Pal and K.~Ray, \emph{{Conformal integrals in various dimensions and
  Clifford groups}},  \href{https://arxiv.org/abs/2303.17326}{{\ttfamily
  2303.17326}}.

\bibitem{Rigatos:2022eos}
K.C.~Rigatos and X.~Zhou, \emph{{Yangian Symmetry in Holographic Correlators}},
  \href{https://doi.org/10.1103/PhysRevLett.129.101601}{\emph{Phys. Rev. Lett.}
  {\bfseries 129} (2022) 101601}
  [\href{https://arxiv.org/abs/2206.07924}{{\ttfamily 2206.07924}}].

\bibitem{Loebbert:2016cdm}
F.~Loebbert, \emph{{Lectures on Yangian Symmetry}},
  \href{https://doi.org/10.1088/1751-8113/49/32/323002}{\emph{J. Phys. A}
  {\bfseries 49} (2016) 323002}
  [\href{https://arxiv.org/abs/1606.02947}{{\ttfamily 1606.02947}}].

\bibitem{Henn:2023tbo}
J.~Henn, E.~Pratt, A.-L.~Sattelberger and S.~Zoia, \emph{{D-module techniques
  for solving differential equations in the context of Feynman integrals}},
  \href{https://doi.org/10.1007/s11005-024-01835-7}{\emph{Lett. Math. Phys.}
  {\bfseries 114} (2024) 87}
  [\href{https://arxiv.org/abs/2303.11105}{{\ttfamily 2303.11105}}].

\bibitem{saito2013grobner}
M.~Saito, B.~Sturmfels and N.~Takayama, \emph{Gr{\"o}bner deformations of
  hypergeometric differential equations}, vol.~6, Springer Science \& Business
  Media (2013).

\bibitem{Ananthanarayan:2022ntm}
B.~Ananthanarayan, S.~Banik, S.~Bera and S.~Datta, \emph{{FeynGKZ: A
  Mathematica package for solving Feynman integrals using GKZ hypergeometric
  systems}}, \href{https://doi.org/10.1016/j.cpc.2023.108699}{\emph{Comput.
  Phys. Commun.} {\bfseries 287} (2023) 108699}
  [\href{https://arxiv.org/abs/2211.01285}{{\ttfamily 2211.01285}}].

\bibitem{Stienstra:2005nr}
J.~Stienstra, \emph{{GKZ hypergeometric structures}},  in \emph{{Instanbul
  2005: CIMPA Summer School on Arithmetic and Geometry Around Hypergeometric
  Functions}}, 11, 2005 [\href{https://arxiv.org/abs/math/0511351}{{\ttfamily
  math/0511351}}].

\bibitem{sattelberger2019d}
A.-L.~Sattelberger and B.~Sturmfels, \emph{D-modules and holonomic functions},
  {\emph{arXiv preprint arXiv:1910.01395} (2019) }.

\bibitem{gel1996hypergeometric}
I.M.~Gel'fand, M.I.~Graev and S.A.~Spirin, \emph{Hypergeometric functions and
  the newton polytope associated with the action of the torus $(c^{*})$ on
  $\bigwedge^k c^{n}$},  in \emph{Doklady Akademii Nauk}, vol.~348,
  pp.~155--158, Russian Academy of Sciences, 1996.

\bibitem{gel1991hypergeometric}
I.M.~Gel'fand, M.I.~Graev and V.S.~Retakh, \emph{Hypergeometric functions on
  the k th exterior power of the space c\^{}n and the grassmannian g\_k,n and
  the connection between them},  in \emph{Doklady Akademii Nauk}, vol.~320,
  pp.~20--24, Russian Academy of Sciences, 1991.

\bibitem{Lee:2013hzt}
R.N.~Lee and A.A.~Pomeransky, \emph{{Critical points and number of master
  integrals}}, \href{https://doi.org/10.1007/JHEP11(2013)165}{\emph{JHEP}
  {\bfseries 11} (2013) 165} [\href{https://arxiv.org/abs/1308.6676}{{\ttfamily
  1308.6676}}].

\bibitem{Cavaglia:2019pow}
A.~Cavagli\`a, N.~Gromov and F.~Levkovich-Maslyuk, \emph{{Separation of
  variables and scalar products at any rank}},
  \href{https://doi.org/10.1007/JHEP09(2019)052}{\emph{JHEP} {\bfseries 09}
  (2019) 052} [\href{https://arxiv.org/abs/1907.03788}{{\ttfamily
  1907.03788}}].

\bibitem{Gromov:2019wmz}
N.~Gromov, F.~Levkovich-Maslyuk, P.~Ryan and D.~Volin, \emph{{Dual Separated
  Variables and Scalar Products}},
  \href{https://doi.org/10.1016/j.physletb.2020.135494}{\emph{Phys. Lett. B}
  {\bfseries 806} (2020) 135494}
  [\href{https://arxiv.org/abs/1910.13442}{{\ttfamily 1910.13442}}].

\bibitem{Maillet:2020ykb}
J.M.~Maillet, G.~Niccoli and L.~Vignoli, \emph{{On Scalar Products in Higher
  Rank Quantum Separation of Variables}},
  \href{https://doi.org/10.21468/SciPostPhys.9.6.086}{\emph{SciPost Phys.}
  {\bfseries 9} (2020) 086} [\href{https://arxiv.org/abs/2003.04281}{{\ttfamily
  2003.04281}}].

\bibitem{Gromov:2022waj}
N.~Gromov, N.~Primi and P.~Ryan, \emph{{Form-factors and complete basis of
  observables via separation of variables for higher rank spin chains}},
  \href{https://doi.org/10.1007/JHEP11(2022)039}{\emph{JHEP} {\bfseries 11}
  (2022) 039} [\href{https://arxiv.org/abs/2202.01591}{{\ttfamily
  2202.01591}}].

\bibitem{Cavaglia:2021mft}
A.~Cavagli\`a, N.~Gromov and F.~Levkovich-Maslyuk, \emph{{Separation of
  variables in AdS/CFT: functional approach for the fishnet CFT}},
  \href{https://doi.org/10.1007/JHEP06(2021)131}{\emph{JHEP} {\bfseries 06}
  (2021) 131} [\href{https://arxiv.org/abs/2103.15800}{{\ttfamily
  2103.15800}}].

\bibitem{Cavaglia:2018lxi}
A.~Cavagli\`a, N.~Gromov and F.~Levkovich-Maslyuk, \emph{{Quantum spectral
  curve and structure constants in $ \mathcal{N}=4 $ SYM: cusps in the ladder
  limit}}, \href{https://doi.org/10.1007/JHEP10(2018)060}{\emph{JHEP}
  {\bfseries 10} (2018) 060}
  [\href{https://arxiv.org/abs/1802.04237}{{\ttfamily 1802.04237}}].

\bibitem{Gelfond:2008ur}
O.A.~Gelfond and M.A.~Vasiliev, \emph{{Higher Spin Fields in Siegel Space,
  Currents and Theta Functions}},
  \href{https://doi.org/10.1088/1126-6708/2009/03/125}{\emph{JHEP} {\bfseries
  03} (2009) 125} [\href{https://arxiv.org/abs/0801.2191}{{\ttfamily
  0801.2191}}].

\bibitem{Vasiliev:2001dc}
M.A.~Vasiliev, \emph{{Relativity, causality, locality, quantization and duality
  in the S(p)(2M) invariant generalized space-time}},
  \href{https://arxiv.org/abs/hep-th/0111119}{{\ttfamily hep-th/0111119}}.

\bibitem{appell1926fonctions}
P.~Appell, \emph{Fonctions Hyperg{\'e}om{\'e}triques et Hypersph{\'e}riques
  Polynomes D’Hermite}, Gauthier-Villars (1926).

\end{thebibliography}\endgroup
\end{document}